%% file: EXO-22-016_temp.tex
\pdfoutput=1
\documentclass[11pt,twoside,a4paper,cmspaper,final,collab]{cms-tdr}
\def\svnVersion{dd89212}\def\svnDate{2023/10/10}\def\cmsCernNoTag{CERN-EP-2023-122}\def\cmsCernDate{\today}\def\cmsMessage{Published in the Journal of High Energy Physics as \href{http://dx.doi.org/10.1007/JHEP10(2023)043}{\doi{10.1007/JHEP10(2023)043}.}}
\begin{document}\cmsNoteHeader{EXO-22-016}

\newcommand{\Eq}[1]{Eq.~(\ref{#1})}
\newcommand{\lumisixteen}{36.3\fbinv\xspace}
\newcommand{\lumiseventeen}{41.5\fbinv\xspace}
\newcommand{\lumieighteen}{59.8\fbinv\xspace}
\newcommand{\lumiT}{138\fbinv\xspace}
\newcommand{\cPm}{\ensuremath{\Pgm}\xspace}
\newcommand{\X}{\ensuremath{{\HepParticle{X}{}{}}}\xspace}
\newcommand{\W}{\ensuremath{{\PW}}\xspace}
\newcommand{\WW}{\ensuremath{\W\W}\xspace}
\newcommand{\WZ}{\ensuremath{\W\PZ}\xspace}
\newcommand{\ZZ}{\ensuremath{\PZ\PZ}\xspace}
\newcommand{\Zprime}{{\HepParticle{Z}{}{\prime}}\xspace}
\newcommand{\tZXtxt}{\ensuremath{\PQt\PZ}+\ensuremath{\X}\xspace}
\newcommand{\tWXtxt}{\ensuremath{\PQt\PW}+\ensuremath{\X}\xspace}
\newcommand{\ttH}{\ensuremath{\PQt\PAQt\PH}\xspace}
\newcommand{\ttW}{\ensuremath{\PQt\PAQt\PW}\xspace}
\newcommand{\ttZ}{\ensuremath{\PQt\PAQt\PZ}\xspace}
\newcommand{\ttGs}{\ensuremath{\PQt\PAQt\PGg^{*}}\xspace}
\newcommand{\ttV}{\ensuremath{\PQt\PAQt\PV}\xspace}
\newcommand{\mmumu}{\ensuremath{m_{\cPm\cPm}}\xspace}
\newcommand{\mbarmumu}{\ensuremath{\overline{m}_{\cPm\cPm}}\xspace}
\newcommand{\mlb}{\ensuremath{m_{\cPm\PQb}}\xspace}
\newcommand{\mlbNs}{\ensuremath{m_{\cPm\PQb\!}}}
\newcommand{\minmlb}{\ensuremath{\min(\mlbNs)}\xspace}
\newcommand{\mZp}{\ensuremath{m_{\Zprime}}\xspace}
\newcommand{\Nb}{\ensuremath{N_{\PQb}}\xspace}
\newcommand{\gZp}{\ensuremath{g_{\Zprime}}\xspace}
\newcommand{\ttwothree}{\ensuremath{\theta_{23}}\xspace}
\newcommand{\gl}{\ensuremath{g_{\ell}}\xspace}
\newcommand{\gb}{\ensuremath{g_{\PQb}}\xspace}
\newcommand{\pileup}{pileup\xspace}
\newcommand{\DrellYan}{Drell--Yan\xspace}

\cmsNoteHeader{EXO-22-016}
\title{Search for a high-mass dimuon resonance produced in association with \PQb quark jets at \texorpdfstring{$\sqrt{s}=13\TeV$}{sqrt(s) = 13 TeV}}

\date{\today}

\abstract{
A search for high-mass dimuon resonance production in association with one or more $\PQb$ quark jets is presented. The study uses proton-proton collision data collected with the CMS detector at the LHC corresponding to an integrated luminosity of $\lumiT$ at a center-of-mass energy of $13\TeV$. Model-independent limits are derived on the number of signal events with exactly one or more than one $\PQb$ quark jet. Results are also interpreted in a lepton-flavor-universal model with $\Zprime$ boson couplings to a $\PQb\PQb$ quark pair ($g_{\PQb}$), an $\PQs\PQb$ quark pair ($g_{\PQb}\delta_{\PQb\PQs}$), and any same-flavor charged lepton ($\gl$) or neutrino pair ($g_{\PGn}$), with $\abs{g_{\PGn}}=\abs{\gl}$. For a $\Zprime$ boson with a mass $\mZp=350\GeV$ (2\TeV) and $\abs{\delta_{\PQb\PQs}} < 0.25$, the majority of the parameter space with $0.0057 < \abs{\gl} < 0.35$ ($0.25 < \abs{\gl} < 0.43$) and $0.0079 < \abs{g_{\PQb}} < 0.46$ ($0.34 < \abs{g_{\PQb}} < 0.57$) is excluded at $95\%$ confidence level. Finally, constraints are set on a $\Zprime$ model with parameters consistent with low-energy $\PQb\to\PQs\ell\ell$ measurements. In this scenario, most of the allowed parameter space is excluded for a $\Zprime$ boson with $350<\mZp<500\GeV$, while the constraints are less stringent for higher $\mZp$ hypotheses. This is the first dedicated search at the LHC for a high-mass dimuon resonance produced in association with multiple $\PQb$ quark jets, and the constraints obtained on models with this signature are the most stringent to date.
}

\hypersetup{
pdfauthor={CMS Collaboration},
pdftitle={Search for a high mass dimuon resonance produced in association with b quark jets at sqr(s)=13 TeV},
pdfsubject={CMS},
pdfkeywords={dimuon, b-tag, Zprime, anomaly, anomalies}
}

\maketitle 

\section{Introduction}
\label{sec:intro}

\begin{figure}[htb!]
  \centering 
  \includegraphics[width=0.35\textwidth]{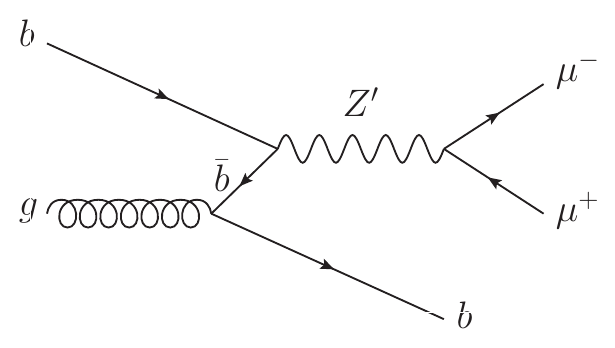}
  \includegraphics[width=0.35\textwidth]{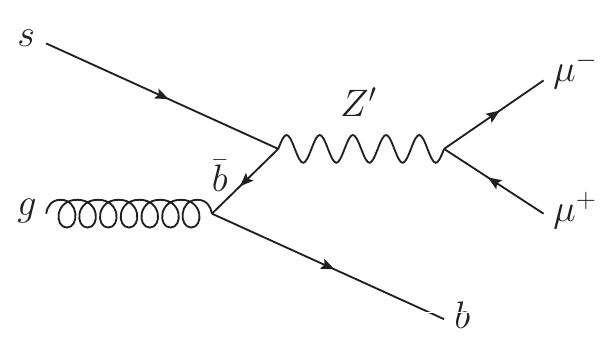}
  \includegraphics[width=0.35\textwidth]{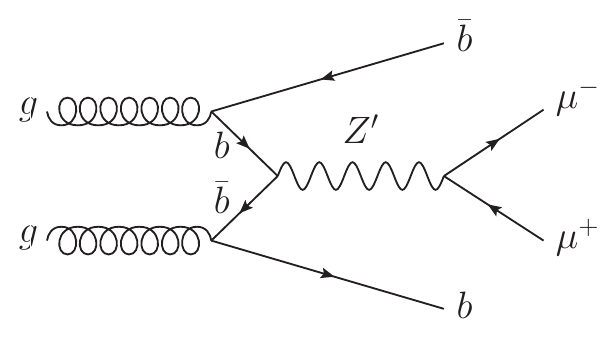}
  \includegraphics[width=0.35\textwidth]{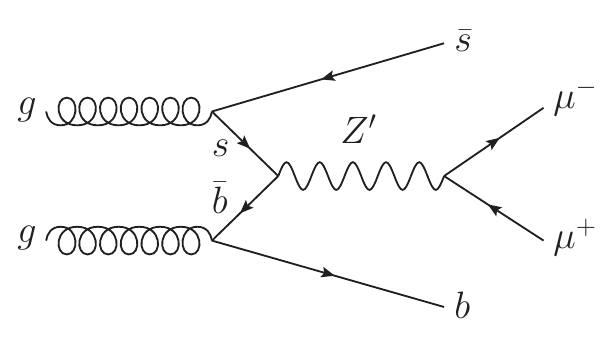}
  \caption{
    Feynman diagrams of $\Zprime \to \Pgmm \Pgmp$ with a $\Zprime$ boson
    produced via $\PQb \PAQb \to \Zprime$ 
    or $\PQs \PAQb \to \Zprime$, with at least one $\PQb$ quark in the final state.
    While a $\Zprime\PQb\PQb$ coupling may be present in any generic model, 
    a $\Zprime\PQs\PQb$ coupling could arise through flavor mixing between the second- and third-generation quarks.
  }
  \label{fig:feynzprime}
\end{figure}

Over the last few years,
a number of experimental results~\cite{LHCb:2013ghj,LHCb:2014cxe,LHCb:2015wdu,LHCb:2015svh,CMS:2015bcy,LHCb:2017avl,Bobeth:2017vxj,CMS:2017rzx,ATLAS:2018gqc,LHCb:2019hip,LHCb:2021trn,LHCb:2021lvy,Parrott:2022rgu}
have suggested that physics beyond the standard model (SM)
could manifest itself in $\PQb \to \PQs \ell^{-} \ell^{+}$ transitions.
Experimental results for this process could be explained by the existence of a new neutral vector boson (\Zprime)
coupling to lepton ($\ell$) pairs~\cite{Leike:1998wr,Langacker:2008yv}
with a mass near the \TeVns scale.
In such a scenario, the new \Zprime boson would couple to \PQb and \PQs quarks,
as represented by the Feynman diagrams in Fig.~\ref{fig:feynzprime}.
While the most recent experimental results~\cite{LHCb:2022qnv,LHCb:2022zom} on the ratio of the branching fractions for
$\PQb \to \PQs \PGmm \PGmp$ to $\PQb \to \PQs \Pem \Pep$ are in agreement with the SM,
there remains some tension in other $\PQb \to \PQs \ell^{-} \ell^{+}$ observables,
such as the overall event rates~\cite{LHCb:2014cxe,LHCb:2015wdu,Parrott:2022rgu} and angular distributions~\cite{LHCb:2013ghj,LHCb:2015svh,CMS:2015bcy,Bobeth:2017vxj,CMS:2017rzx,ATLAS:2018gqc},
although theoretical predictions for these are less well established than those for the ratio of branching fractions.
The current situation was recently discussed in Ref.~\cite{Allanach:2022iod}.

Inclusive searches for beyond-the-SM (BSM) \Zprime bosons
have already been performed at the CERN LHC~\cite{ATLAS:2019erb,EXO-19-019}.
However, they are limited by a large \DrellYan (DY) background
and might not be sensitive to scenarios in which the \Zprime boson couples preferentially to second- or third-generation quarks.
Therefore, the goal of this analysis is to search for a $\Zprime \to \PGmm \PGmp$ resonance
with an explicit requirement on the presence of $\PQb$ quark jets,
which strongly disfavors DY events.

The relevant interactions can be described through a lepton-flavor-universal (LFU) model with a Lagrangian simplified from the one in Ref.~\cite{ParticleDataGroup:2022pth},
\begin{multline}
\label{eq:LFUlagrangian}
\mathcal{L}_\text{BSM} = \Zprime_{\eta}~\biggl\{
\gl \sum_{\ell=\Pe, \Pgm, \Pgt}{\HepAntiParticle{\ell}{}{}\gamma^{\eta}P_{\text{L}}\ell}
+
g_{\PGn} \sum_{\PGn=\PGne, \PGnGm, \PGnGt}{\PAGn\gamma^{\eta}P_{\text{L}}\PGn}\\
+
\gb \left[\PAQb \gamma^{\eta}P_{\text{L}}\PQb + \delta_{\PQb\PQs} \left( \PAQs\gamma^{\eta} P_{\text{L}}\PQb + ~\text{h.c.} \right) \right]
\biggr\},
\end{multline}
where $P_{\text{L}}$ denotes the left-handed projection operator.
There are four coupling parameters in this model:
a common $\gl$ coupling for all charged leptons,
a common $g_{\PGn}$ coupling for all neutrinos,
the $g_{\PQb}$ coupling that scales both $\Zprime\PQb\PQb$ and $\Zprime\PQs\PQb$ interactions,
and the separate $\delta_{\PQb\PQs}$ parameter that solely scales the $\Zprime\PQs\PQb$ interaction.
In this analysis, we probe values of $\delta_{\PQb\PQs}$ between 0 and 0.25: The $\Zprime\PQs\PQb$ interaction is fully suppressed at $\delta_{\PQb\PQs}=0$,
and $\delta_{\PQb\PQs}$ values above $0.25$ may yield scenarios incompatible with the measurement of the mass difference between the mass eigenstates of the neutral \PBs mesons~\cite{ParticleDataGroup:2022pth}.

The $B_{3}\!-\!L_{2}$ model is a less generic $\Zprime$ model introduced in 2017~\cite{Alonso:2017uky,Bonilla:2017lsq}
to accommodate the experimental $\PQb \to \PQs \ell^{-} \ell^{+}$ anomalies.
It is based on a new $B_{3}\!-\!L_{2}~U(1)$ gauge symmetry in which $B_3$ and $L_2$ represent the third-generation baryon and muon lepton numbers, respectively.
In this model, \gZp denotes the coupling of the \Zprime boson to SM fermions,
and the angle \ttwothree controls the mixing angle between the second- and third-generation quarks.
This mixing angle is equivalent to $\arctan(-\delta_{\PQb\PQs})$ in the notation of \Eq{eq:LFUlagrangian}.
The allowed parameter space for the $B_{3}\!-\!L_{2}$ model has been recently reevaluated in Ref.~\cite{Allanach:2022iod}.

Studies of dilepton invariant mass distributions in final states with $\PQb$ quark jets
have previously been performed at the LHC~\cite{ATLAS:2019odt,CMS:2019mij,ATLAS:2021mla}.
However, these studies either focused on final states with exactly one $\PQb$ quark jet, or used inclusive samples with at least one such jet,
or searched for nonresonant signatures in the high dimuon invariant mass tail.
Some of the previous studies were based on data sets with only about one fourth of the integrated luminosity used in this analysis.
Furthermore, they all suffered from large backgrounds arising from the pair production of $\PQt$ quarks (\ttbar),
each decaying to one $\PQb$ quark and one $\PW$ boson, with $\PW\to\ell\PGn$.
In this analysis, we reduce this background substantially, as noted below, and we further optimize the event categorization 
for the presence of one or more $\PQb$ quark jets.
We focus on dimuon resonances
because the experimental tensions that motivate this search are thought
to be consistent with BSM contributions in the muon sector~\cite{Allanach:2022iod}.

The analysis can be summarized as follows:
\begin{itemize}
\item The search is performed for a narrow $\Zprime \to \PGmm\PGmp$ resonance with a mass $\mZp>350\GeV$ 
in the presence of at least one $\PQb$ quark jet.
The width of the resonance is assumed to be narrow relative to the dimuon invariant mass ($\mmumu$) resolution.
The validity of this assumption is ensured by restricting the search to the regions of parameter space where
the $\Zprime$ width does not exceed one half of the $\mmumu$ resolution.
The resolution itself ranges from $6.6\GeV$ at $\mZp=350\GeV$ to $30\,(106)\GeV$ at $\mZp=1\,(2.5)\TeV$.
\item The data sample is collected with the CMS detector during the LHC Run~2 data-taking period of 2016--2018
at a center-of-mass energy of 13\TeV, and corresponds to an integrated luminosity of $\lumiT$ 
(2016: $\lumisixteen$; 2017: $\lumiseventeen$; 2018: $\lumieighteen$).
\item Events in this data sample are categorized according to the multiplicity of $\PQb$ quark jets: $\Nb = 1$ and $\Nb \geq 2$. 
Tight and relaxed requirements are used together in identifying these jets, as discussed in Section~\ref{sec:CMS},
in order to maximize the sensitivity to a possible signal.
Events with $\Nb = 0$ are not considered because the background is about 150 times larger in this category
due to the contribution from DY events.
\item The dominant backgrounds across the probed \mmumu range arise from the DY process and \ttbar production.
The DY background is already reduced by requiring the detection of at least one $\PQb$ quark jet.
The \ttbar background is suppressed by requiring that the minimum invariant mass of any muon--$\PQb$ quark 
jet pairing ($\minmlb$) be $>\!175\GeV$, \ie, above the $\PQt$ quark mass.
Such a requirement reduces the \ttbar background by more than two orders of magnitude
while retaining most of the predicted signal with large $\mZp$,
thus offering an improved signal discrimination compared to other kinematic requirements used in previous similar studies.
Other background sources, including \tZXtxt, \tWXtxt, $\ttV$ ($\PV = \PW$, $\PZ$, $\PGg^{*}$), and $\ttH$ processes, 
as well as diboson ($\WW$, $\WZ$, and $\ZZ$) production,
are less important and further reduced by vetoing events with any additional lepton or isolated high transverse momentum (\pt)
charged hadron. 
The veto for events with isolated charged hadrons is used to enhance the rejection of hadronically decaying $\PGt$ leptons and 
unidentified electrons or muons.
\item We extract constraints on the total event yield 
as a function of the fraction of events in each of the $\Nb = 1$ and $\Nb \geq 2$ categories.
The constraints are determined using unbinned maximum likelihood fits of
the \mmumu distributions with analytic functions within mass ranges sliding coherently with the probed value of \mZp,
as further discussed in Section~\ref{sec:analysis}.
The functional forms used in the fit and the general statistical procedure are also described in Section~\ref{sec:analysis}
and are similar to those in Ref.~\cite{CMS:2021sch}.
The constraints on the total event yield are then reinterpreted in terms of the parameters of 
specific models with a \Zprime boson coupled to $\PQb$ quarks.
\end{itemize}

All of the results for this analysis can also be found on HEPData~\cite{hepdata}.

\section{Experimental setup}
\label{sec:CMS}

The central feature of the CMS apparatus is a superconducting solenoid of 6\unit{m} internal diameter, providing a magnetic field of 3.8\unit{T}. Within the solenoid volume are a silicon pixel and strip tracker (up to a pseudorapidity coverage of $\abs{\eta}<2.5$), a lead tungstate crystal electromagnetic calorimeter ($\abs{\eta}<3$), and a brass and scintillator hadron calorimeter (HCAL, with $\abs{\eta}<3$), each composed of a barrel and two endcap sections. 
Forward calorimeters extend the $\eta$ coverage provided by the barrel and endcap detectors up to $\abs{\eta}=5$. 
Muons are measured in gas-ionization detectors ($\abs{\eta}<2.4$) embedded in the steel flux-return yoke outside the solenoid.
A more detailed description of the CMS detector, together with a definition of the coordinate system used 
and the relevant kinematic variables, can be found in Ref.~\cite{Chatrchyan:2008zzk}.

Events of interest are selected using a two-tiered trigger system. The first level, composed of custom hardware processors, uses information from the calorimeters and muon detectors to select events at a rate of around 100\unit{kHz} within a fixed latency of about 4\mus~\cite{CMS:2020cmk}. The second level, known as the high-level trigger (HLT), consists of a farm of processors running a version of the full event reconstruction software optimized for fast processing, and reduces the event rate to around 1\unit{kHz} before data storage~\cite{CMS:2016ngn}.

A particle-flow (PF) algorithm~\cite{Sirunyan:2017ulk} aims 
to reconstruct and identify each individual particle in an event, 
using an optimized combination of all subdetector information.
The particles reconstructed with this algorithm are hereafter referred to as PF candidates,
and their isolation is measured from the flux of other charged PF candidates 
within a cone of $\Delta R=\sqrt{\smash[b]{(\Delta \phi)^2 + (\Delta \eta)^2}}=0.3$ around their direction,
where $\phi$ is the azimuthal angle.

The primary proton-proton ($\Pp\Pp$) interaction vertex (PV) is taken to be the vertex corresponding to the hardest scattering in the event, evaluated using tracking information alone (as described in Section~9.4.1 of Ref.~\cite{CMS-TDR-15-02}).
Muons originating from the PV
are identified using a set of requirements~\cite{Sirunyan:2018fpa} 
that are optimized to maximize their selection efficiency at high $\pt$,
\ie, roughly above $200\GeV$.
The muon \pt assignment is performed with the `TuneP' algorithm~\cite{CMS:2019ied},
which mainly relies on the tracker information instead of data from the PF algorithm.
This avoids the potential bias arising from electromagnetic showers in the muon detectors,
caused by extra particles radiated by the high-\pt muon.
The difference between the \pt value assigned using only tracker information and that obtained from the PF algorithm
is taken into account when applying requirements on other PF-based quantities in the events.
As the analysis targets high-\pt muons, which may radiate photons when traversing detector material, 
muons are required to be isolated with respect to nearby charged-particle tracks, \ie, the presence of neutral PF candidates is not considered in assessing the level of isolation.
The tracker-only isolation of muons,
which is defined as the scalar \pt sum of all tracks, excluding that of the muon, 
within a longitudinal distance of $0.2\cm$ from the PV
and a cone of $\Delta R=0.3$ relative to the muon track, 
is required to be less than 5\GeV and less than 5\% of the muon \pt as measured in the tracker.

Events are collected with triggers that require the presence of at least one muon at HLT with
$\pt > 50\GeV$ and $\abs{\eta} < 2.4$.
The trigger selection efficiency is measured using the ``tag-and-probe" method~\cite{CMS:2010svw} and independent data sets.
The efficiency varies from about 70\% in the detector endcaps, \ie, at $\abs{\eta}\gtrsim2.1$,
to about 95\% in the detector barrel, yielding an overall efficiency of about 90\% with only
a slight dependence on the data-taking period~\cite{CMS:2021yvr}.

Events are further required to contain
two oppositely charged muons with $\pt > 53\GeV$ and $\abs{\eta} < 2.4$ satisfying the
identification and isolation criteria,
with an overall muon reconstruction and selection efficiency larger than $90\%$.
The two muons are then used
to form a $\PGmm\PGmp$ resonance candidate.
Events with additional muons of $\pt > 10\GeV$ and $\abs{\eta}<2.4$
satisfying the same identification and isolation criteria are vetoed.
We also veto events that contain isolated PF muon candidates
originating from the PV with $\pt > 5\GeV$.
These vetoes, together with those described below, 
reduce the SM background arising from \tZXtxt, \tWXtxt, \ttV, \ttH,
and diboson production.

In the analysis, electrons are identified using loose requirements~\cite{CMS:2020uim}, 
for the purpose of vetoing events with electrons originating from the PV.
The electron isolation is measured from the flux of PF photons and hadron candidates
within a cone of $\Delta R=0.3$ around its direction, 
including corrections for contributions from additional $\Pp\Pp$ interactions 
within the same or nearby bunch crossings (\pileup).
This PF-based isolation is required to be below a threshold that varies as a function of the electron \pt.
Events with any such electron of $\pt > 10\GeV$ and $\abs{\eta}<2.5$ are vetoed,
together with events that contain any isolated PF electron candidate originating from the PV with $\pt > 5\GeV$.

The largest fraction of all visible \PGt lepton decay products consists of a single charged hadron,
often accompanied by multiple neutral pions.
For this reason,
the veto for events with extra leptons is expanded to include 
isolated PF charged-hadron candidates originating from the PV with $\pt > 10\GeV$.

\begin{figure}[htb!]
  \centering
    \includegraphics[width=0.55\textwidth]{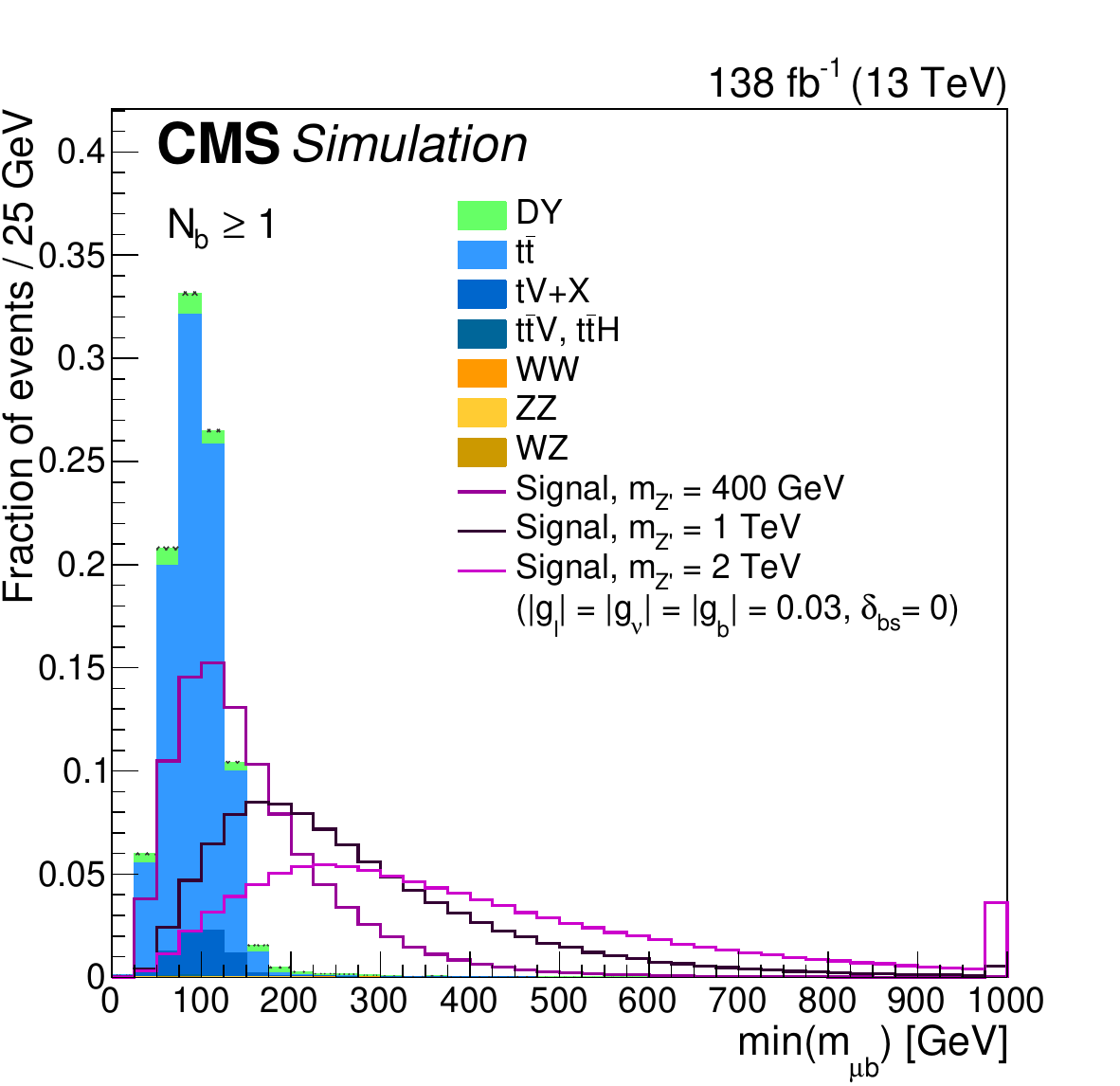}
    \caption{
    Distribution of \minmlb as obtained from simulation in events with $\Nb\geq1$
    passing all the other selection requirements. In this search, we require $\minmlb>175\GeV$.
    The stacked histogram displays the expected distribution from the simulation of the SM backgrounds, while 
    the overlaid open histograms illustrate the size and shape of the \Zprime contribution 
    from the LFU model described in \Eq{eq:LFUlagrangian},
    for several \Zprime mass hypotheses.
    For illustrative purposes, we choose couplings $\abs{\gl} = \abs{g_{\PGn}} = \abs{\gb} = 0.03$ and $\delta_{\PQb\PQs}=0$.
    The efficiency of the requirement on the \Zprime signal process is about 30\%, 70\%, and 85\% for $\mZp=400$, $1000$, and $2000\GeV$, respectively.
    The contribution of background processes other than DY and \ttbar is so small that it is only barely visible at the bottom of the stacked histogram.
    The hatched region indicates the statistical uncertainty arising from the limited size of the SM simulated samples (Section~\ref{sec:MC}).
    Histograms are normalized to unit area.
    }
    \label{fig:minmlb}
\end{figure}

Jets are reconstructed using the anti-$\kt$ algorithm~\cite{Cacciari:2008gp,Cacciari:2011ma} with
a distance parameter of 0.4.
Jet energies are corrected for instrumental effects and contributions from \pileup{}~\cite{Sirunyan:2020foa}.
We select jets with $\pt > 20\GeV$ and $\abs{\eta}<2.5$, that must also be
separated by $\Delta R>0.4$ from each muon candidate.
The identification of $\PQb$ quark jets uses the $\textsc{DeepJet}$ algorithm~\cite{Bols:2020bkb,CMS-DP-2023-005}. 
At least one $\PQb$ quark jet must satisfy tight identification criteria 
($\sim$58\% tagging efficiency for $\sim$0.1\% misidentification probability of jets from light quarks and gluons), 
and all other $\PQb$ quark jets are identified using relaxed requirements 
($\sim$76\% tagging efficiency for $\sim$1\% misidentification probability).
In order to reject $\PGm\PQb$ pairs that originate from $\PQt$ quark decays, 
we calculate all possible values of \mlb among the selected muons and $\PQb$ quark jets, and require $\minmlb>175\GeV$, \ie,  
above the value of the $\PQt$ quark mass.
This requirement is used to suppress the $\ttbar$ background, which is reduced by more than two orders of magnitude,
while retaining most of the possible signals with large \mZp,
as shown in Fig.~\ref{fig:minmlb}.

The missing transverse momentum vector (\ptvecmiss) is estimated
from the negative of the vector \pt sum of all PF candidates,
where the pileup-per-particle-identification algorithm~\cite{Bertolini:2014bba,Sirunyan:2020foa} weights each PF candidate
for its probability to originate from the PV in order to reduce the \pileup dependence of this observable.
The \ptvecmiss vector is corrected further for residual inconsistencies that may be introduced from the non-PF high-\pt muon identification criteria and the muon \pt assignment through the `TuneP' algorithm.
While no source of significant genuine \ptmiss is expected except for neutrinos in the case of semileptonic decays 
of the heavy-flavor quarks, large \ptmiss may nevertheless be observed if the momentum of a muon or $\PQb$ quark jet is mismeasured.
Thus, events are vetoed if $\ptmiss > 250\GeV$ and \ptvecmiss
is aligned ($\abs{\Delta\phi} < 0.3$) or antialigned ($\abs{\Delta\phi} > \pi-0.3$)
with any selected muon or $\PQb$ quark jet.
This requirement retains almost the totality of the possible signals.
Other anomalous high-\ptmiss events, due to reconstruction issues, detector malfunctions, or noncollision backgrounds, are rejected by dedicated filters.
These filters identify more than 85--90\% of these spurious high-\ptmiss events with a mistagging rate less than 0.1\%~\cite{CMS:2019ctu}.
In addition, the three-dimensional angle between the two muons is required to be smaller than $\pi-0.02$ 
to further suppress cosmic ray muon contributions.

In 2018, a sector of the HCAL ($-3.2<\eta<-1.3$ and $-1.57<\phi<-0.87$) was not operational 
for a data-taking period corresponding to an integrated luminosity of $39\fbinv$,
resulting in the misreconstruction of hadronic jets in that $\eta$--$\phi$ range.
Therefore, events collected during this period are rejected if jets or electrons are found in the affected region,
with upper and lower boundaries for jets enlarged by $0.2$ in both $\eta$ and $\phi$ in order to account for the size of the jet cone.
The simulation of events, discussed in Section~\ref{sec:MC}, is also adjusted by applying the same requirement in a subset of events corresponding to the fraction of data affected by this issue.
The efficiency loss over the full data set is about 5\% regardless of the value of $\mmumu$.

\section{Event simulation}
\label{sec:MC}

We use simulated samples of various BSM $\Zprime$ models to motivate the event selection
and provide model-dependent limits.
On the other hand, 
the SM background is estimated directly from data as a continuum background 
in the \mmumu spectrum, parametrized by analytic functions.
Thus, the background estimation does not rely on simulation.
The SM simulated samples are used only to validate
and optimize the event selection as well as to
visually compare the observed dimuon mass distribution to the expected SM background from simulation.

The simulated samples for the dominant DY and \ttbar backgrounds, and the \tWXtxt process,
are generated using the \POWHEG~2~\cite{Frixione:2007vw,Alioli:2009je,Melia:2011tj,Nason:2013ydw} program 
at next-to-leading order (NLO) in perturbative quantum chromodynamics (QCD).
Samples for the \tZXtxt, \ttW, \ttZ, and \ttGs processes, and for the \ttH process are generated at NLO in QCD 
using \MGvATNLO~\cite{Alwall:2014hca}~v2.6.5 and v2.6.1, respectively,
with the FxFx~\cite{Frederix:2012ps} scheme to match jets from matrix element calculations and parton showers.
Samples for diboson production are generated using either \MGvATNLO~v2.6.5 or \POWHEG~2 at NLO in QCD.

As discussed in Section~\ref{sec:intro},
we consider a signal \Zprime model involving
$\Zprime\PQb\PQb$ and $\Zprime\PQs\PQb$ couplings and $\Zprime\to\PGm\PGm$ decays.
We generate such interactions
using \MGvATNLO~v2.9.9 at LO in QCD
with \mZp ranging from $350$ to $2500\GeV$.
For all the generated samples, the total width $\Gamma(\Zprime)$ of the \Zprime
is always smaller than one half of the \mmumu resolution
to ensure the validity of the narrow width approximation intrinsic to the search strategy.
Predictions for any other model with a likewise narrow width value can be obtained through
reweighting by the double ratio of the quantity
$\Gamma(\Zprime \to \PQq_{\text{i}} \PAQq_{\text{j}}) \Gamma(\Zprime \to \Pgmm \Pgmp)/\Gamma(\Zprime)$
between the target model and the generated sample.
Here, $\Gamma(\Zprime \to \PQq_{\text{i}} \PAQq_{\text{j}})$ denotes the partial width for
$\Zprime\to\PQb\PAQb$, $\Zprime\to\PQs\PAQb$~and~$\Zprime\to\PQb\PAQs$, 
and $\Gamma(\Zprime \to \Pgmm \Pgmp)$ is the partial width for $\Zprime\to\PGmm\PGmp$.
The reweighting is performed by separating the simulated event sample into
contributions from matrix elements with different
$\PQq_{\text{i}} \PAQq_{\text{j}} \to \Zprime$ interactions, as shown in
Fig.~\ref{fig:feynzprime},
and the use of partial widths ensures generality across different ways to construct target models.

In all samples, the parton shower and hadronization are modeled with \PYTHIA 8.230~\cite{Sjostrand:2014zea}
using the \textsc{CP5}~\cite{Sirunyan:2019dfx} tune.
The parton distribution functions are taken from NNPDF~3.1~\cite{NNPDF:2017mvq}.
For the SM simulated samples, they are taken at next-to-NLO in QCD.
For the signal simulation, they are taken at LO in QCD to match the signal cross section calculations.
All the simulated events are generated with a distribution of additional $\Pp\Pp$ interactions per bunch crossing 
that is adjusted to match the corresponding \pileup distribution measured in data.
Finally, the detector response is simulated with the \GEANTfour~\cite{Agostinelli2003250} package.

\section{Analysis strategy}
\label{sec:analysis}

The events selected as described in Section~\ref{sec:CMS} are categorized
according to the multiplicity of $\PQb$ quark jets: $\Nb = 1$ and $\Nb \geq 2$.
The search is performed by simultaneously fitting the unbinned \mmumu distributions in these two categories with analytic functions,
within mass ranges that vary coherently with the value of \mZp to be probed.

We parametrize the signal \mmumu distribution as the sum of a Gaussian distribution 
with a double-sided Crystal Ball function~\cite{Oreglia:1980cs,Gaiser:1982yw}
with varying proportions, and a common resolution parameter $\sigma_\text{mass}$.
The choice of the functional form and the dependence of the corresponding parameters on \mZp are determined from simulation.
The fits are performed using a finite grid of
simulated signal samples, and the fit parameters are subsequently interpolated across \mZp.
As mentioned in Section~\ref{sec:intro}, the values of $\sigma_\text{mass}$ range 
from $6.6\GeV$ at $\mZp=350\GeV$ 
to $30\,(106)\GeV$ at $\mZp=1\,(2.5)\TeV$.
Because of the narrow-width approximation intrinsic to the search strategy,
the values of $\sigma_\text{mass}$ determined from simulation as a function of \mZp are independent of the signal coupling values.
Fits to the data are performed within an \mmumu mass window of ${\pm}10\, \sigma_{\text{mass}}$
around the probed \mZp value
with the restriction $\mmumu>275\GeV$.
The fit window is chosen to be relatively wide 
to ensure that the background is reliably estimated,
without any significant bias that may otherwise arise in the presence of a potential signal at its center.
The window slides coherently with the probed value of \mZp{}, which ranges between $350$ and $2500\GeV$, in steps reflecting the size of $\sigma_\text{mass}$.
This approach is used, instead of fitting the full \mmumu distribution at once,
so as not to rely on prior knowledge of the SM background shape across the full range of interest.

With a relatively wide mass window, the background \mmumu distribution can be modeled analytically using Bernstein polynomials, exponential, or power-law functions.
The choice of the functional form for the background model is treated as a discrete nuisance parameter in the fit.
An envelope of functional forms is provided to the fit. Then, in the minimization process of the negative logarithm of the likelihood, 
the discrete profiling method~\cite{Dauncey:2014xga} is used to select the best fit background model as a function of the signal strength.
The statistical procedures to determine the goodness of fit for the backgrounds and extract the final fit results are similar 
to those described in Ref.~\cite{CMS:2021sch}.
For Bernstein polynomials, the best order in each fit mass window and event category is selected 
by means of a \textit{Fisher test}~\cite{ftest}.
Typically, first order Bernstein polynomials are selected,
since higher order functions do not significantly improve the description of the data
because of the small number of observed events.
For the same reason, only Bernstein polynomials of the first order are used to model the backgrounds in windows with less than ten observed events.
To avoid artifacts from fluctuations due to limited event samples,
models that would yield an increasing background as a function of \mmumu within the fit mass window are rejected.
Only background models that fit the data appropriately are considered: a goodness-of-fit test is performed based on a $\chi^2$ test statistic.
This is converted into a $p$-value through a set of pseudo-experiments, and models with $p<0.01$ are rejected.

The potential presence of a bias in the measurement of a signal due to the choice of the background functional forms is assessed by means of pseudo-experiments.
First, a varying amount of signal is injected on top of the background generated according to
the selected functional form. Then, a background$+$signal fit is performed allowing the signal
yield to float freely. The bias is quantified as the difference between the measured and injected signal yields
relative to the statistical uncertainty in the measured signal yield.
No evidence of a statistically significant bias is found.

A systematic uncertainty of 1.6\% in the expected signal yields 
arises from the integrated luminosity measurement~\cite{CMS-LUM-17-003,CMS-PAS-LUM-17-004,CMS-PAS-LUM-18-002}.
In addition, uncertainties in signal event yields from jet energy scale, trigger, muon reconstruction and $\PQb$ tagging efficiencies, 
and finite simulated sample size are included in the fits.
We include a 5\% uncertainty in the signal acceptance 
to account for possible mismodeling of the muon identification efficiency in the simulation.
We also assess an uncertainty $\lesssim$5\% arising from the estimation of the size of the fit mass window.
Finally, we account for possible effects on the signal shape
from the uncertainties in the muon momentum scale ($\lesssim$0.1\% at $\mZp=1\TeV$) and resolution (${\lesssim}10\%\  \sigma_{\text{mass}}$).
A summary of the sources of signal uncertainty and their sizes is given in Table~\ref{table:syst}.
Uncertainties from other sources are found to be negligible.

\begin{table}[!h]
\centering
\caption{
Summary of signal uncertainties together with their sizes or size ranges.
The uncertainties are grouped based on whether they affect the normalization or the shape of the signal,
and any variations for the two categories of \Nb are shown.
The uncertainties in the signal normalization are expressed as relative uncertainties (\%) with respect to the nominal expected yields.
The fit parameter $\mbarmumu$ defines the position of the maximum of the reconstructed \mmumu distribution for a given \mZp hypothesis.
The fit resolution parameter $\overline{\sigma}_{\text{mass}}$ is distinguished from the values of $\sigma_{\text{mass}}$ extracted from simulation.
}
\begin{tabular}{l c c c}
Source & \multicolumn{2}{c}{Normalization (\%)}  & Shape \\
       & $\Nb=1$ & $\Nb\geq 2$                   &       \\
\hline
Integrated luminosity & \multicolumn{2}{c}{1.6}  & \NA \\
Trigger               & \multicolumn{2}{c}{1--5} & \NA \\
Jet energy scale      & 1--1.5 & 2--5            & \NA \\
$\PQb$ tagging        & 1      & 5               & \NA \\
\PGm reconstruction   & \multicolumn{2}{c}{2.5}  & \NA \\
\PGm identification   & \multicolumn{2}{c}{5}    & \NA \\
Size of fit window & \multicolumn{2}{c}{$\lesssim5$} & \NA \\
Size of simulated sample & $<1$ & $<5$           & \NA \\ 
\hline
\PGm momentum scale in $\mbarmumu$ & \multicolumn{2}{c}{\NA}  & $\lesssim0.1\%\,\mZp^2/\TeVns$ \\
\PGm momentum resolution in $\overline{\sigma}_{\text{mass}}$ & \multicolumn{2}{c}{\NA} & $\lesssim10\%\,\sigma_{\text{mass}}$ \\
\end{tabular}
\label{table:syst}
\end{table}

\section{Results}
\label{sec:results}

\begin{figure}[htb!]
  \centering
    \includegraphics[width=0.45\textwidth]{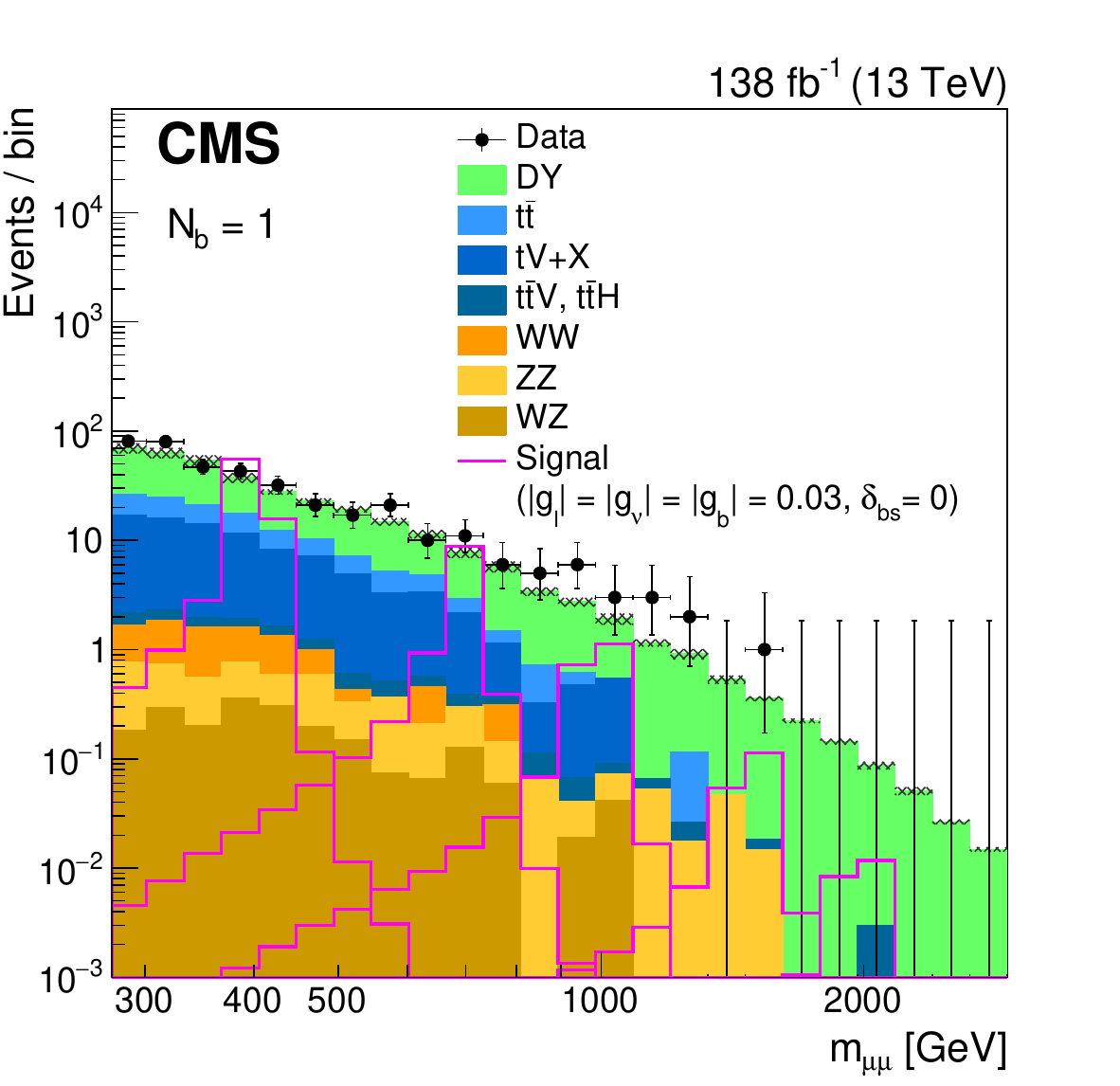}
    \includegraphics[width=0.45\textwidth]{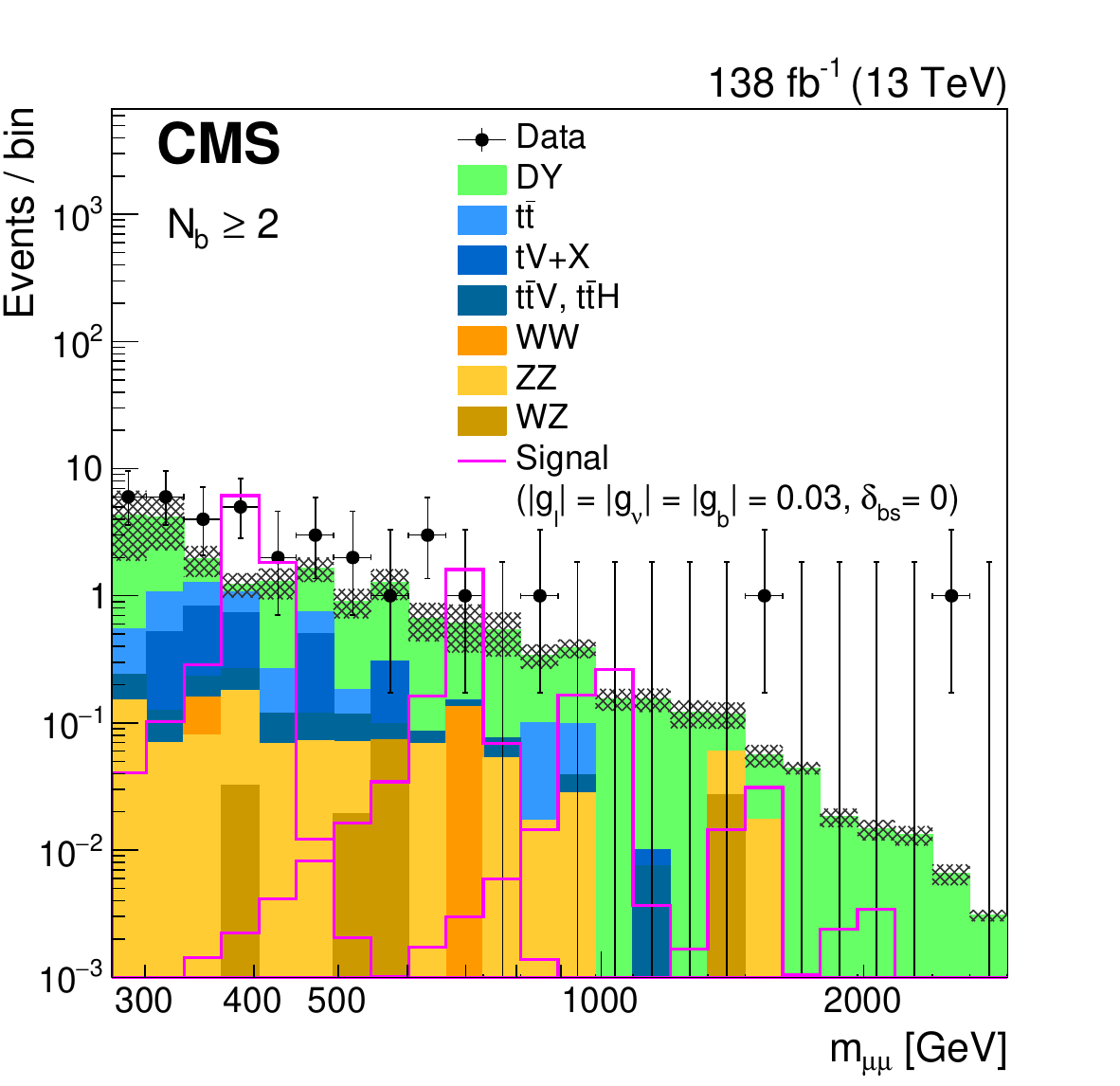}
    \caption{
    Distributions of \mmumu in the $\Nb=1$ (left) and $\Nb\geq2$ (right) event categories.
    The stacked histogram displays the expected distribution from the SM background simulation.
    The overlaid open distributions illustrate the \Zprime contribution from the LFU model
    at $\abs{\gl} = \abs{g_{\PGn}} = \abs{\gb} = 0.03$ and $\delta_{\PQb\PQs}=0$ for a variety of \Zprime mass hypotheses.
    The observed data are shown as black points with statistical error bars.
    The hatched region indicates the statistical uncertainty arising from the limited size of the SM simulated samples.
    The size of the bins increases as a function of \mmumu.
    In extracting the results of the search, the background is estimated directly from data, so the SM background simulation is only illustrative in these distributions.
    }
    \label{fig:dimuonmass}
\end{figure}

The \mmumu distributions in the two event categories are shown in Fig.~\ref{fig:dimuonmass}
for data, SM background simulation, and several representative signal hypotheses. 
As already mentioned, the SM background simulation is only used for illustrative purposes
since the background is estimated directly from data across the full \mmumu range of interest.

\begin{figure}[htb!]
  \centering
  \includegraphics[width=0.45\textwidth]{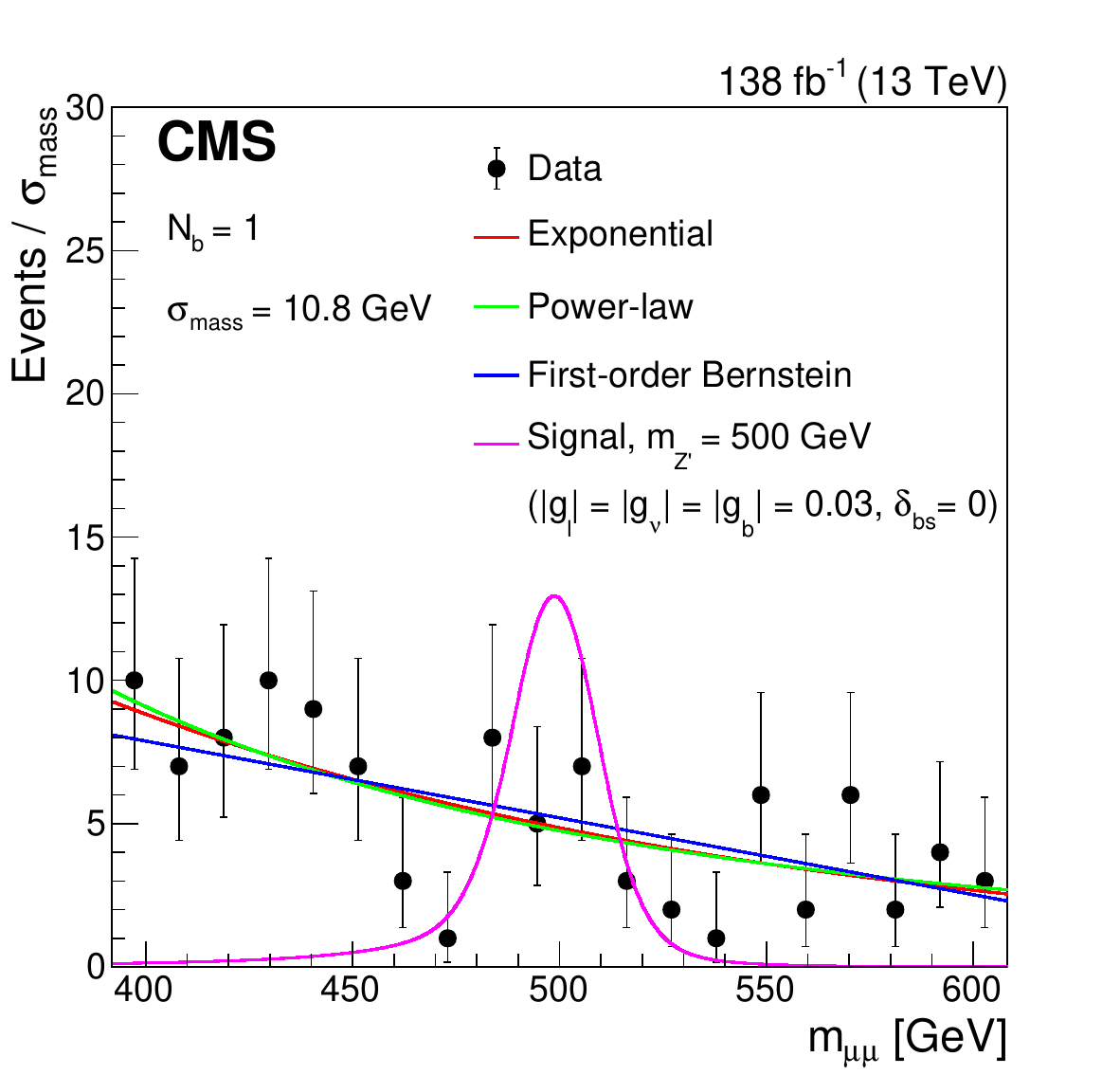}
  \includegraphics[width=0.45\textwidth]{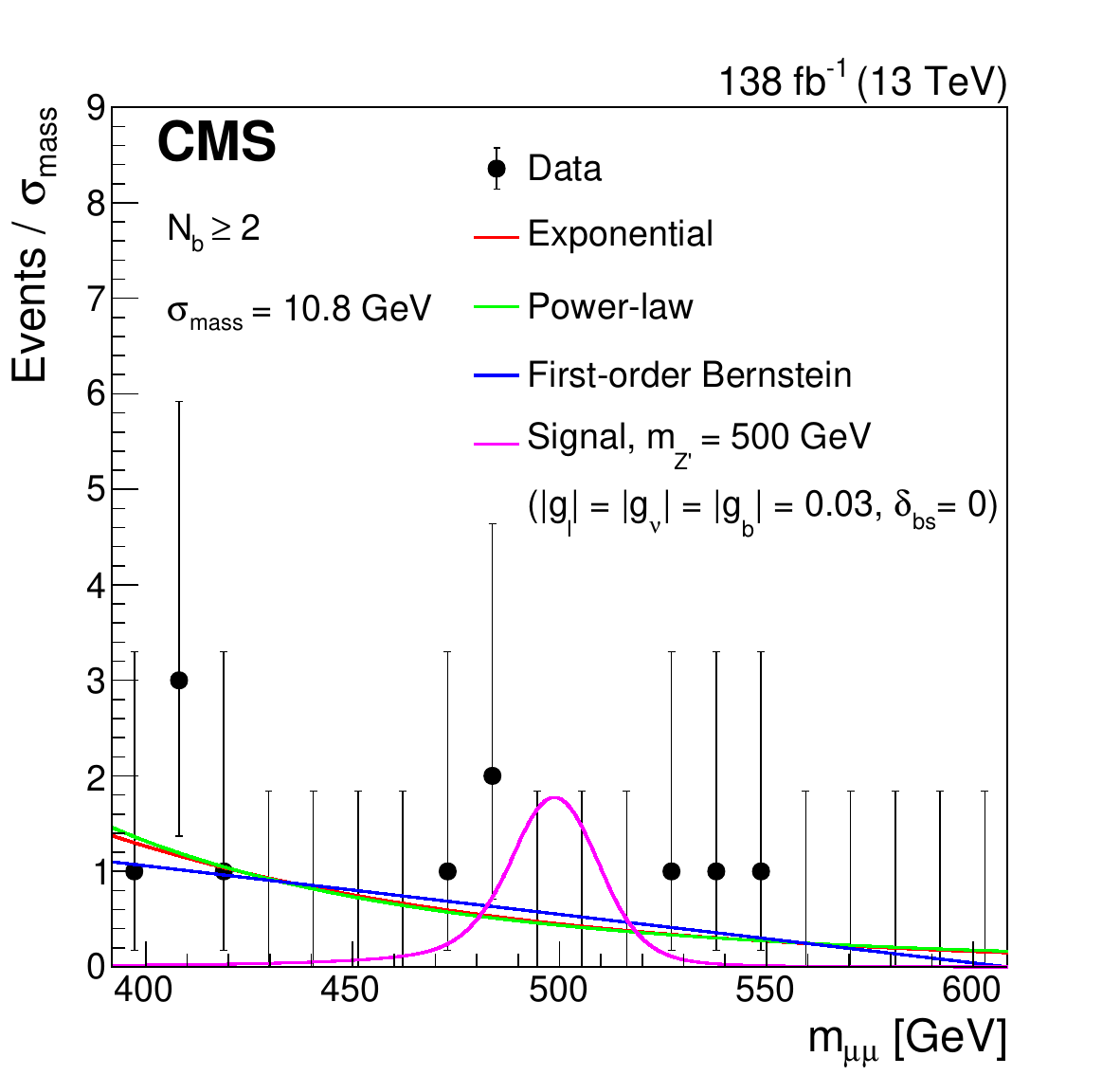}
  \caption{
    Invariant mass \mmumu distributions in the $\Nb=1$ (left) and
    $\Nb\geq2$ (right) categories, shown 
    together with the corresponding selected background functional forms used as input to 
    the \textit{discrete profiling} method~\cite{Dauncey:2014xga} when probing the $\mZp=500\GeV$ hypothesis. 
    The expected signal distribution
    for the LFU model described in \Eq{eq:LFUlagrangian},
    with couplings $\abs{\gl} = \abs{g_{\PGn}} = \abs{\gb} = 0.03$ and $\delta_{\PQb\PQs}=0$, is overlaid.
    The displayed mass range corresponds to the fit window used for this \mZp hypothesis,
    which is ${\pm} 10\, \sigma_{\text{mass}}$ around the probed \mZp value.
    While the likelihood fits are performed on unbinned data, here we present the data in
    binned histograms with binning chosen to reflect the size of $\sigma_{\text{mass}}$.
    }
    \label{fig:fitBG}
\end{figure}

The results of a single set of background-only fit to data in the event categories 
with $\Nb=1$ and $\Nb\geq2$ are shown in Fig.~\ref{fig:fitBG} for a sample signal mass hypothesis, $\mZp=500\GeV$,
within the corresponding \mmumu fit window.
No significant excess is observed over the background-only expectation in any of the mass windows explored.

\begin{figure}[htb!]
  \centering
  \includegraphics[width=0.45\textwidth]{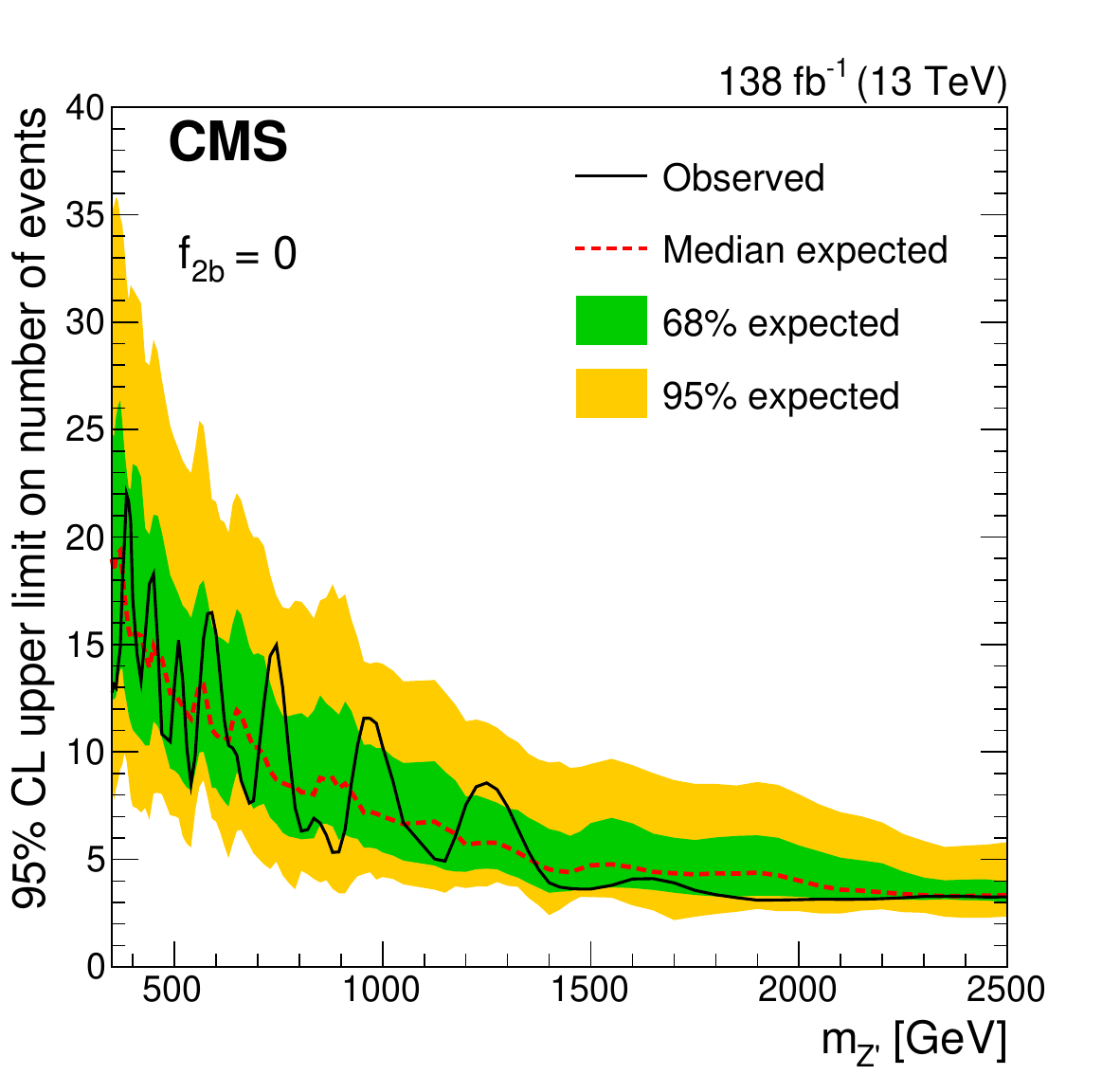}
  \includegraphics[width=0.45\textwidth]{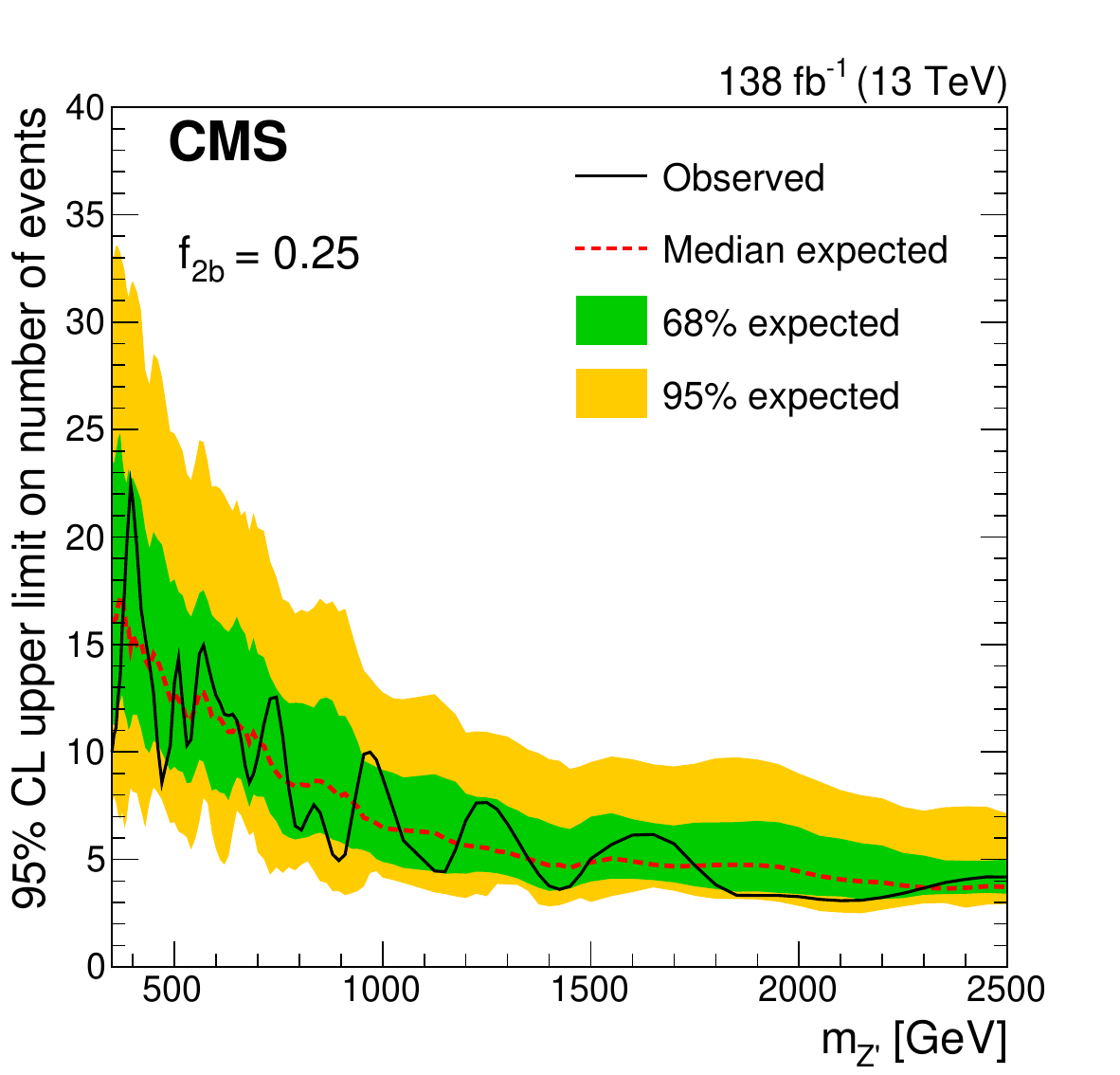}
  \includegraphics[width=0.45\textwidth]{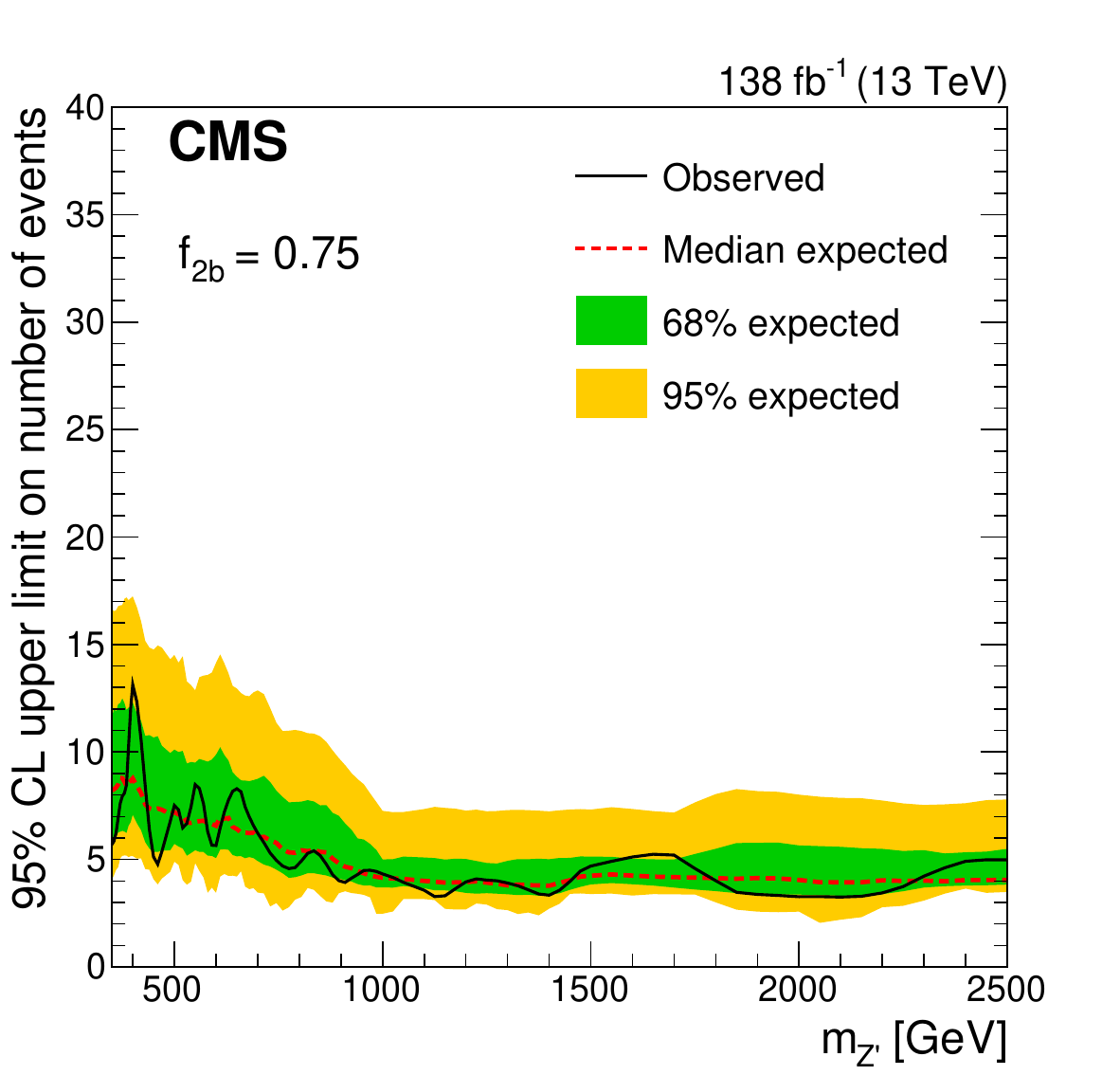}
  \includegraphics[width=0.45\textwidth]{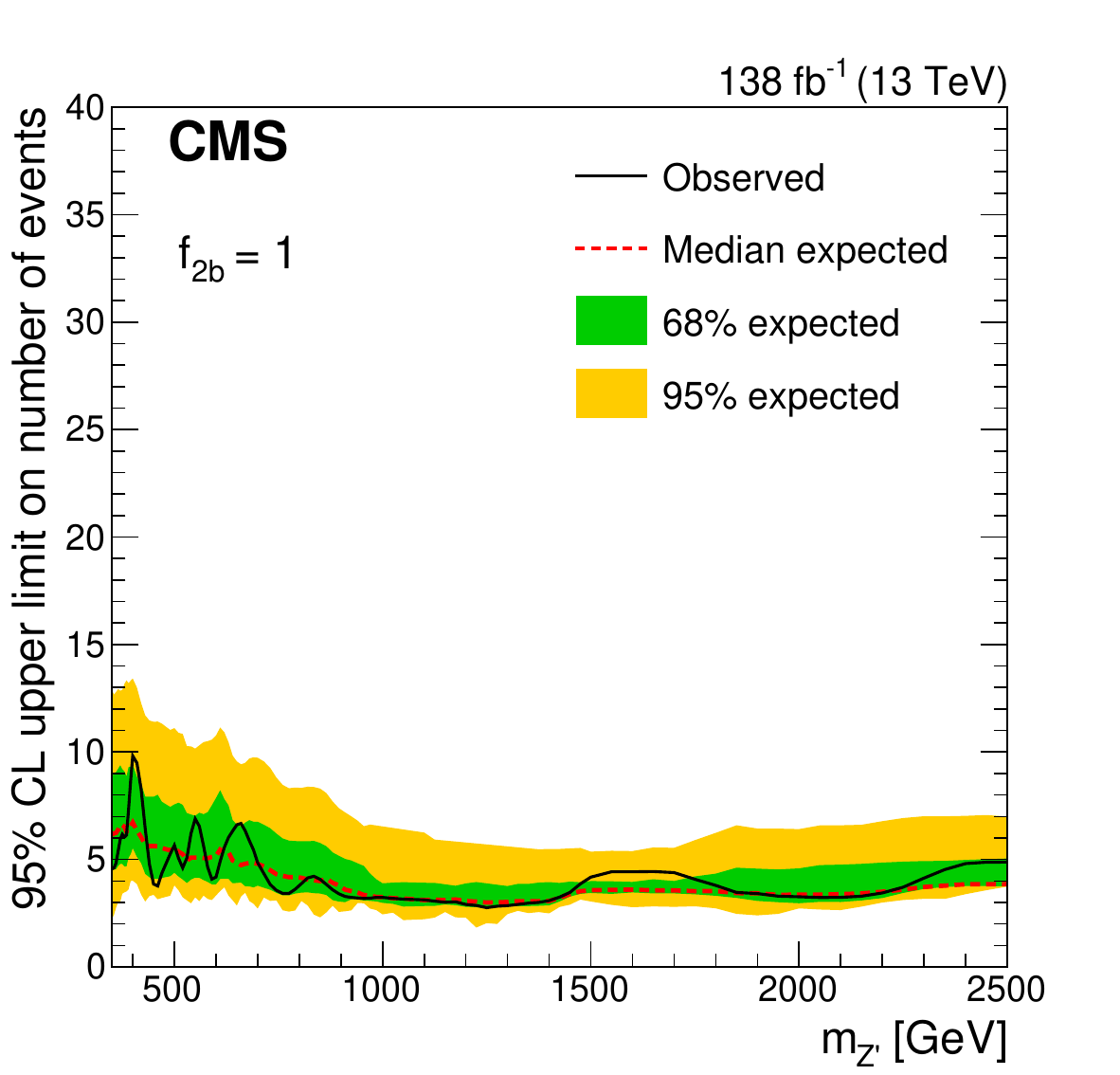}
  \caption{Exclusion limits at 95\%~\CL on the number of selected
    BSM events with $\Nb \geq 1$ as functions of \mZp
    for the different representative values of $f_{2\PQb}=0$ (upper left), 
    $0.25$ (upper right), $0.75$ (lower left),
    and $1$ (lower right).
    The quantity $f_{2\PQb}$ is the fraction of BSM
    events passing the analysis selection that have at least two \PQb quark jets.
    The solid black (dashed red) curve represents the observed (median expected) exclusion.
    The inner green (outer yellow) band indicates the region containing 68 (95)\% 
    of the distribution of limits expected under the background-only hypothesis.
    }
  \label{fig:limits-nomodel}
\end{figure}

The results are used to set model-independent limits at 95\% confidence level (\CL) 
on the number of signal events with $\Nb \geq 1$.
The relative fraction of events in the $\Nb\geq2$ category, $f_{2\PQb}$, is varied
to probe a range of hypotheses of signal production in association with $\PQb$ quarks.
These limits are shown in Fig.~\ref{fig:limits-nomodel}.
They are extracted by performing unbinned fits to the \mmumu distribution in data, using windows that slide in steps reflecting the size of $\sigma_\text{mass}$.
As different data events are included in each of the fit windows, fluctuations are present in both the expected and observed exclusion limits.
The frequency of these fluctuations follows the size of the steps used to probe different \mZp hypotheses,
and their amplitude is typically less than the 68\% expected band.
The statistical procedure to set the limits follows a modified frequentist approach employing the \CLs 
criterion~\cite{Junk:1999kv,Read:2002hq,Cowan:2010js,ATLAS:2011tau}.

\begin{figure}[htb!]
\centering
\includegraphics[width=0.45\textwidth]{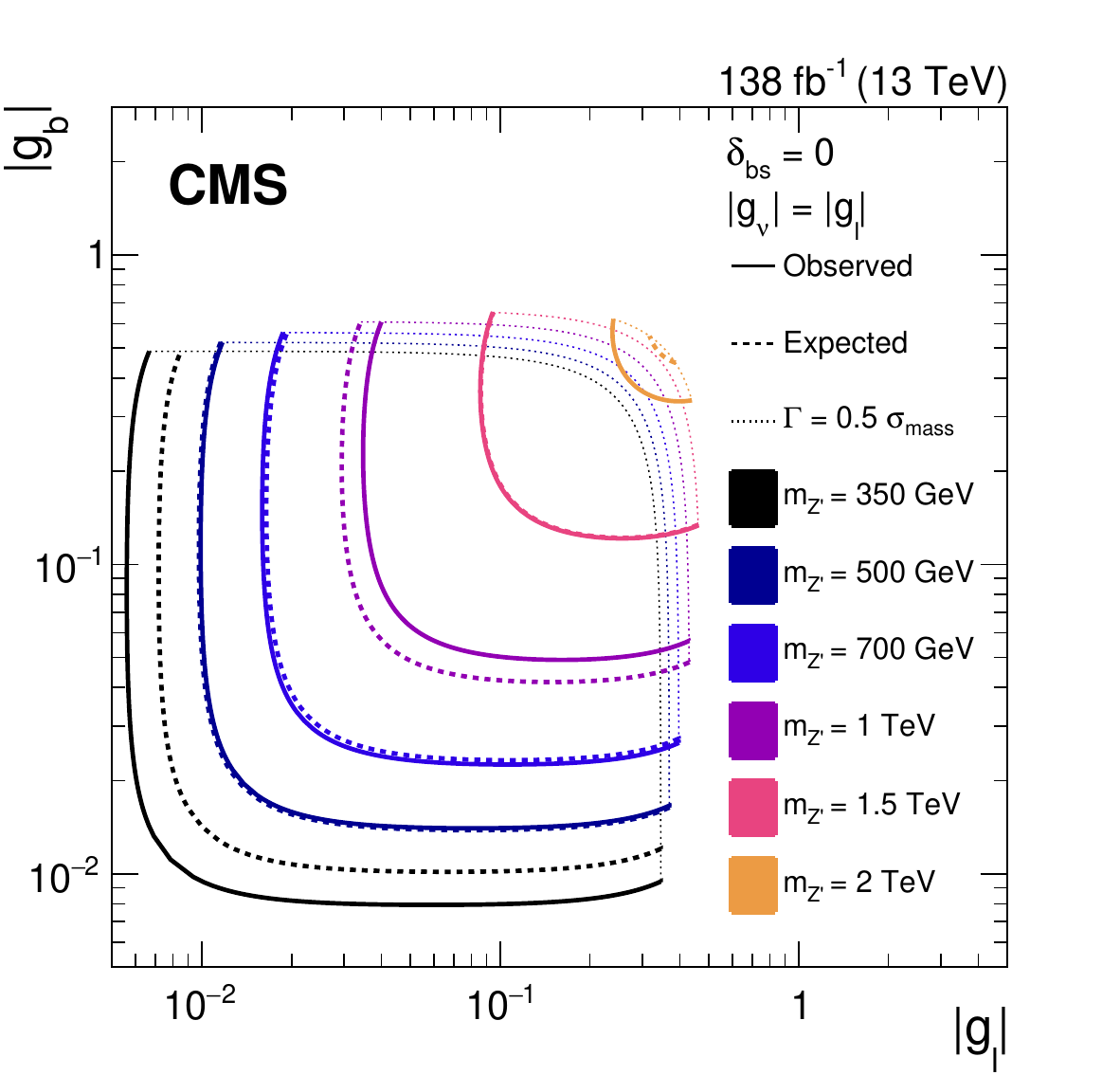}
\includegraphics[width=0.45\textwidth]{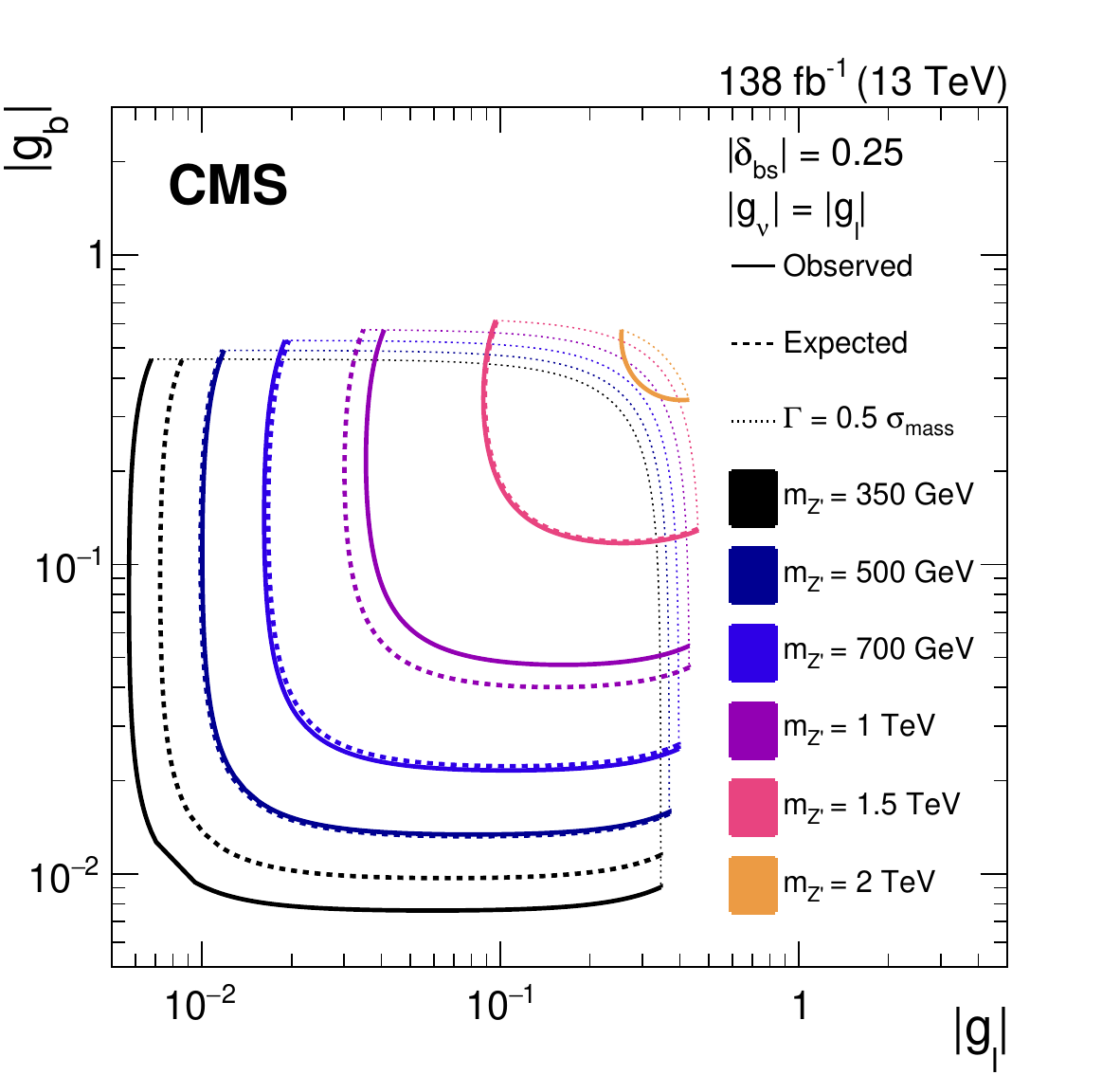}
\caption{
Observed (solid) and median expected (dashed) exclusion limits at 95\%~\CL
in the $\abs{\gb}$--$\abs{\gl}$ plane for the LFU model.
The scenarios considered have $\abs{\delta_{\PQb\PQs}}$ values of either 0 (left) or 0.25 (right).
In all cases, we assume $\abs{g_{\PGn}} = \abs{\gl}$.
The exclusion limits are given up to coupling values at which the $\Zprime$ width is equal to half of
the $\Pgm\Pgm$ invariant mass resolution, marked by the dotted curves.
Beyond these coupling values, the narrow width approximation intrinsic to the search strategy is not considered valid.
The enclosed regions are excluded.
For $\abs{\delta_{\PQb\PQs}}=0.25$ and $\mZp=2\TeV$, no region of the parameter space is expected to be excluded.
}
\label{fig:genericLFU}
\end{figure}

Figure~\ref{fig:genericLFU} provides an interpretation 
for the LFU model described in \Eq{eq:LFUlagrangian}
with the assumed values $\abs{g_{\PGn}}=\abs{\gl}$, and either $\abs{\delta_{\PQb\PQs}}=0$ or $0.25$. 
As $\abs{\gl}$ or $\abs{\gb}$ increases, the
expected value of $\Gamma(\Zprime)$
may approach and exceed the
value of $\sigma_{\text{mass}}$.
Since $\Gamma(\Zprime)$ is assumed to be smaller than $\sigma_{\text{mass}}$ in obtaining these results,
we restrict our exclusion ranges to the regions where $\Gamma(\Zprime) < \sigma_\text{mass} / 2$.
For a $\Zprime$ boson with a mass of $350\GeV$ ($2\TeV$) and $\abs{\delta_{\PQb\PQs}} < 0.25$, 
the majority of the parameter space with $0.0057 < \abs{\gl} < 0.35$ ($0.25 < \abs{\gl} < 0.43$) 
and $0.0079 < \abs{\gb} < 0.46$ ($0.34 < \abs{\gb} < 0.57$) is excluded 
at 95\%~\CL.

\begin{figure}[htb!]
  \centering
  \includegraphics[width=0.45\textwidth]{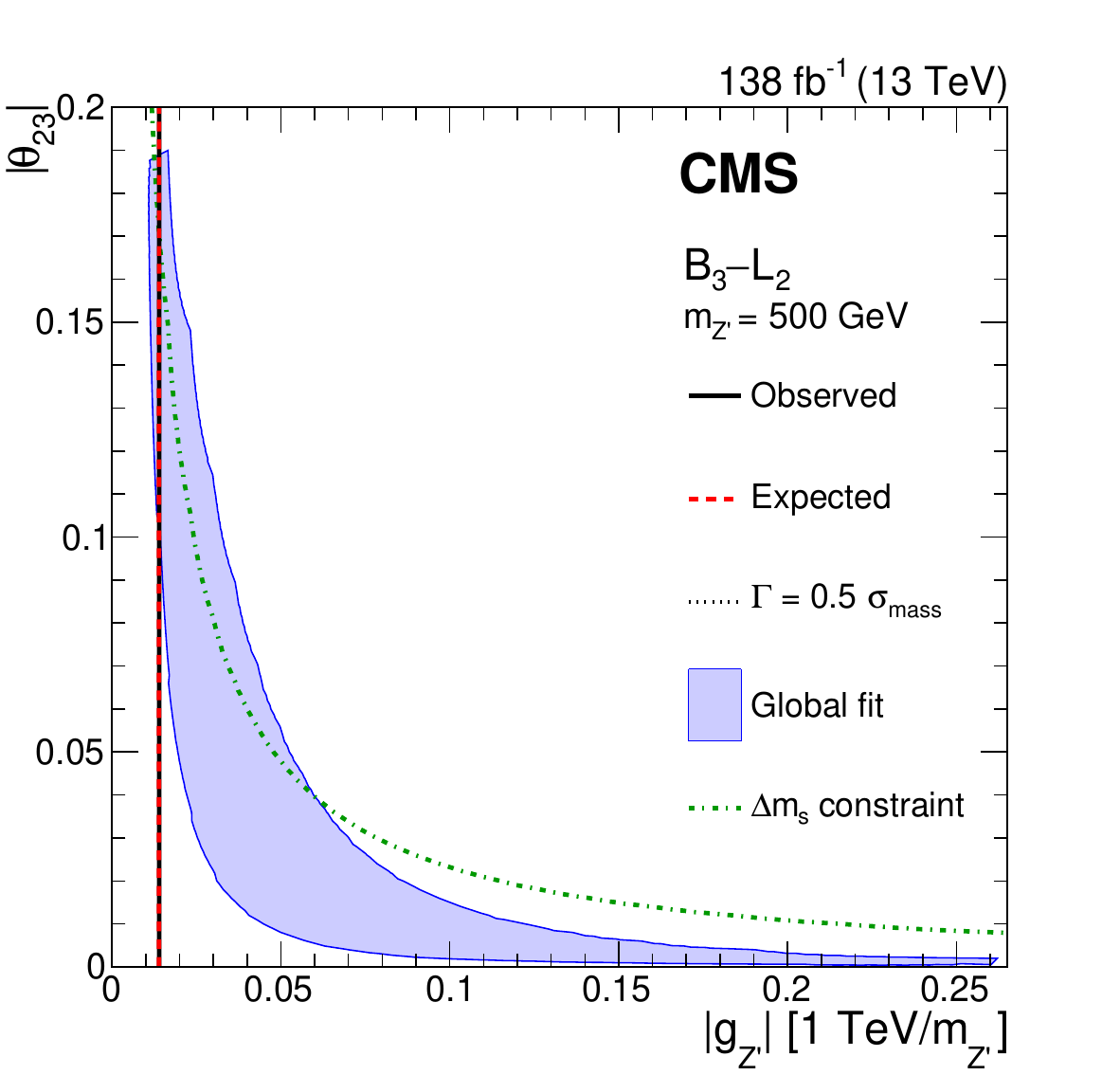}
  \includegraphics[width=0.45\textwidth]{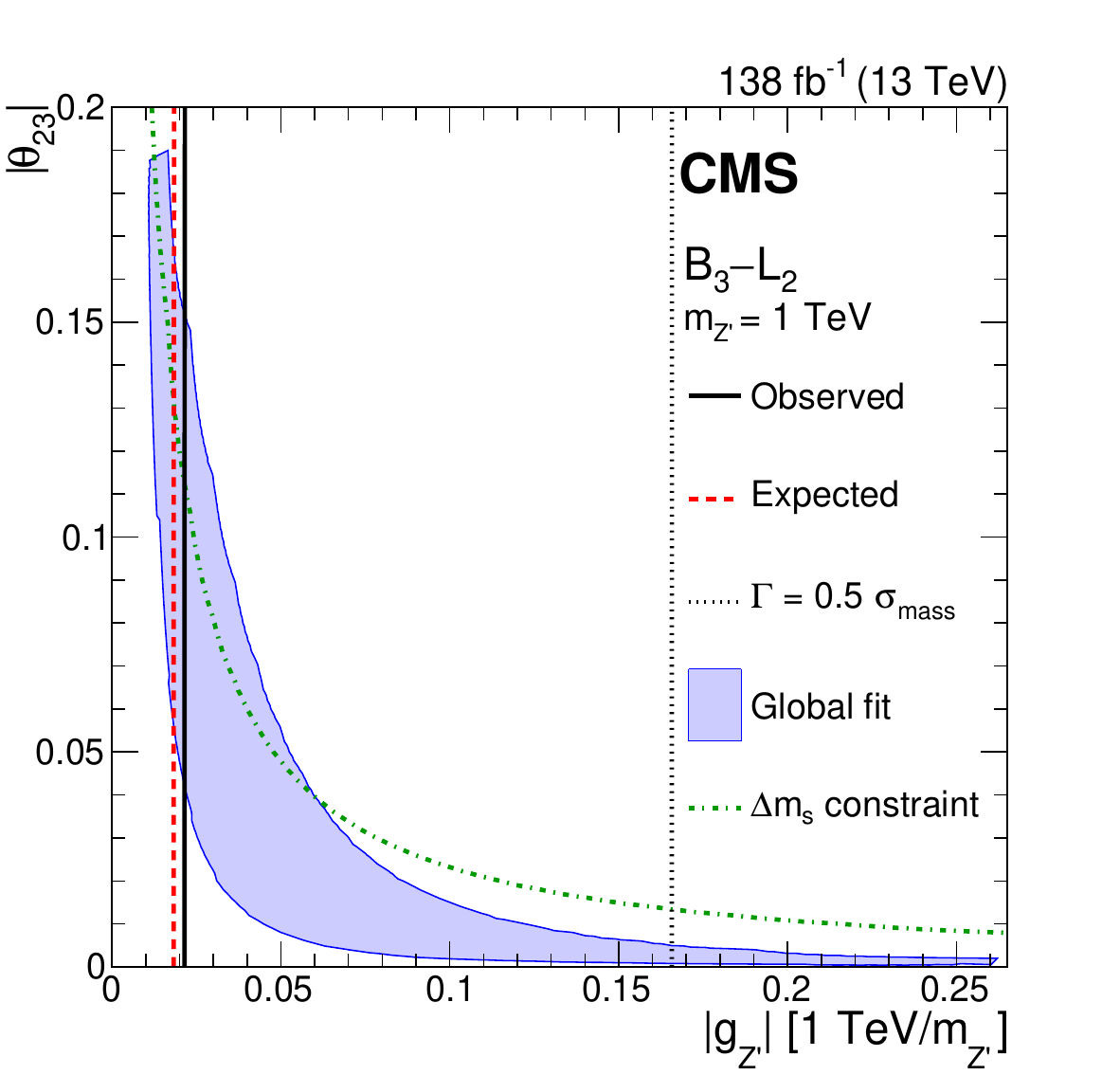}
  \includegraphics[width=0.45\textwidth]{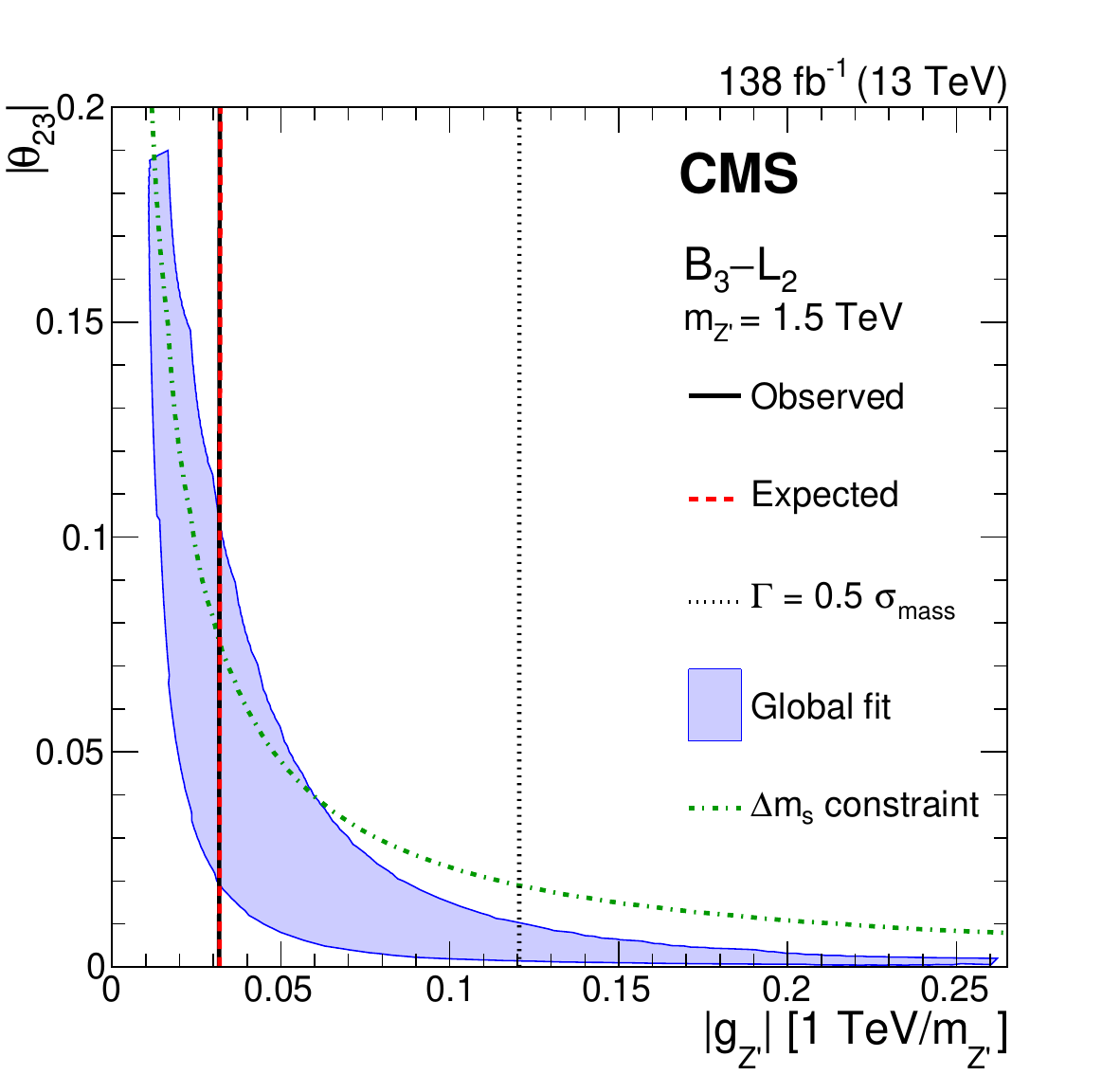}
  \includegraphics[width=0.45\textwidth]{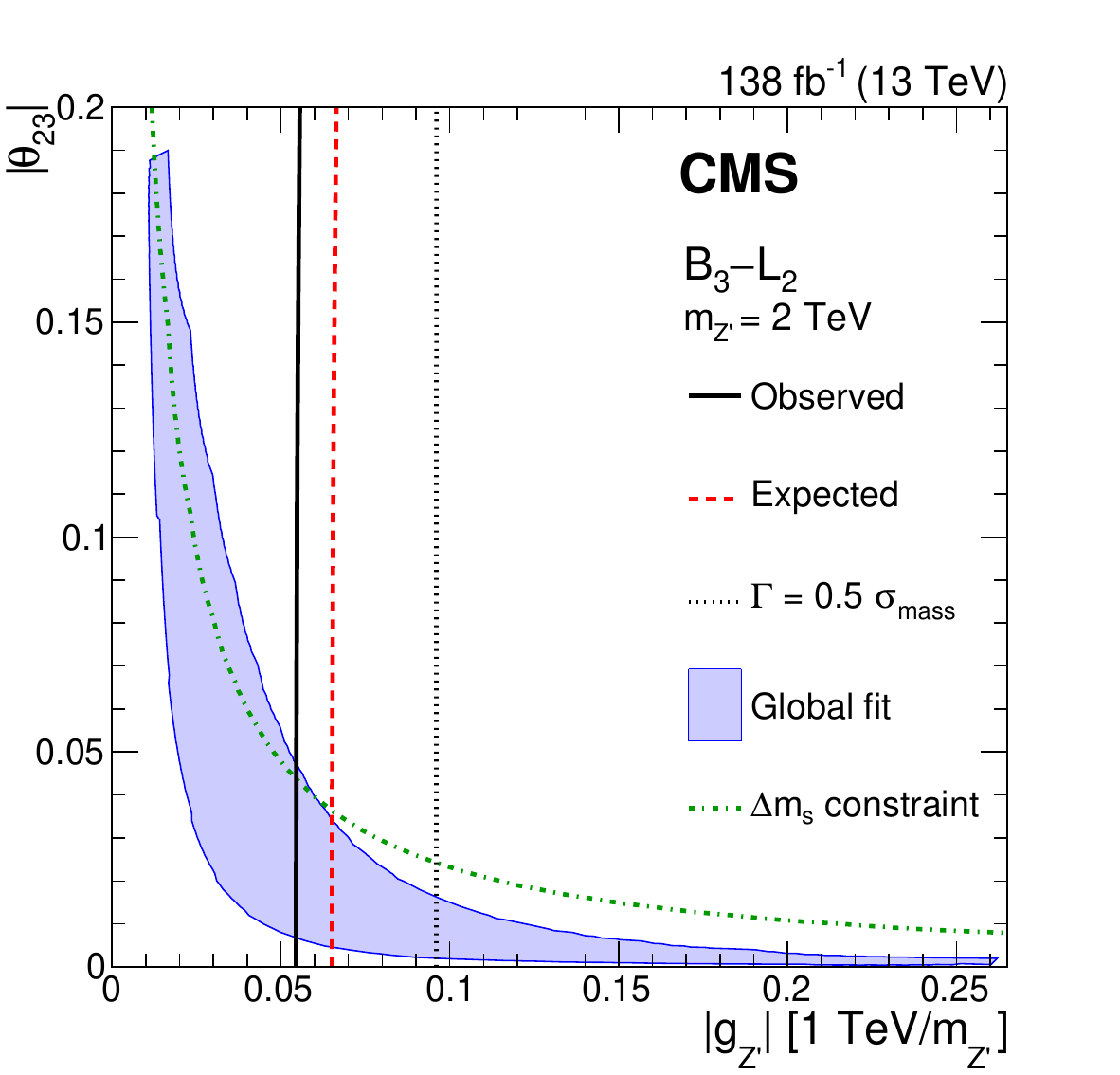}
  \caption{Exclusion limits at 95\%~\CL in the $\abs{\ttwothree}$--$\abs{\gZp}$
    plane for the $B_{3}\!-\!L_{2}$ model~\cite{Allanach:2022iod}, for representative values of
    $\mZp=500\GeV$ (upper left), $\mZp=1\TeV$ (upper right), $\mZp=1.5\TeV$ (lower left), and $\mZp=2\TeV$ (lower right).
    The solid black (dashed red) curves represent the observed (median expected) exclusions.
    The dotted curves denote the coupling values at which the \Zprime width
    equals one half of the $\Pgm\Pgm$ invariant mass resolution.
    For larger values of the couplings, the narrow width approximation intrinsic to the search strategy is not considered valid.
    For a given mass, the region enclosed between the solid black (dashed red) and dotted curves is 
    (expected to be) excluded.
    The dotted curve for $\mZp = 500\GeV$ lies beyond the displayed $\abs{\gZp}$ range and is, therefore, not shown.
    The shaded blue area represents the region preferred from the global fit in Ref.~\cite{Allanach:2022iod} at 95\%~\CL.
    The region above the green dash-dotted curve is incompatible at 95\%~\CL with
    the measurement of the mass difference between the mass eigenstates of the neutral \PBs mesons~\cite{ParticleDataGroup:2022pth}.
  }
  \label{fig:limits_b3ml2}
\end{figure}

\begin{figure}[htb!]
  \centering
  \includegraphics[width=0.45\textwidth]{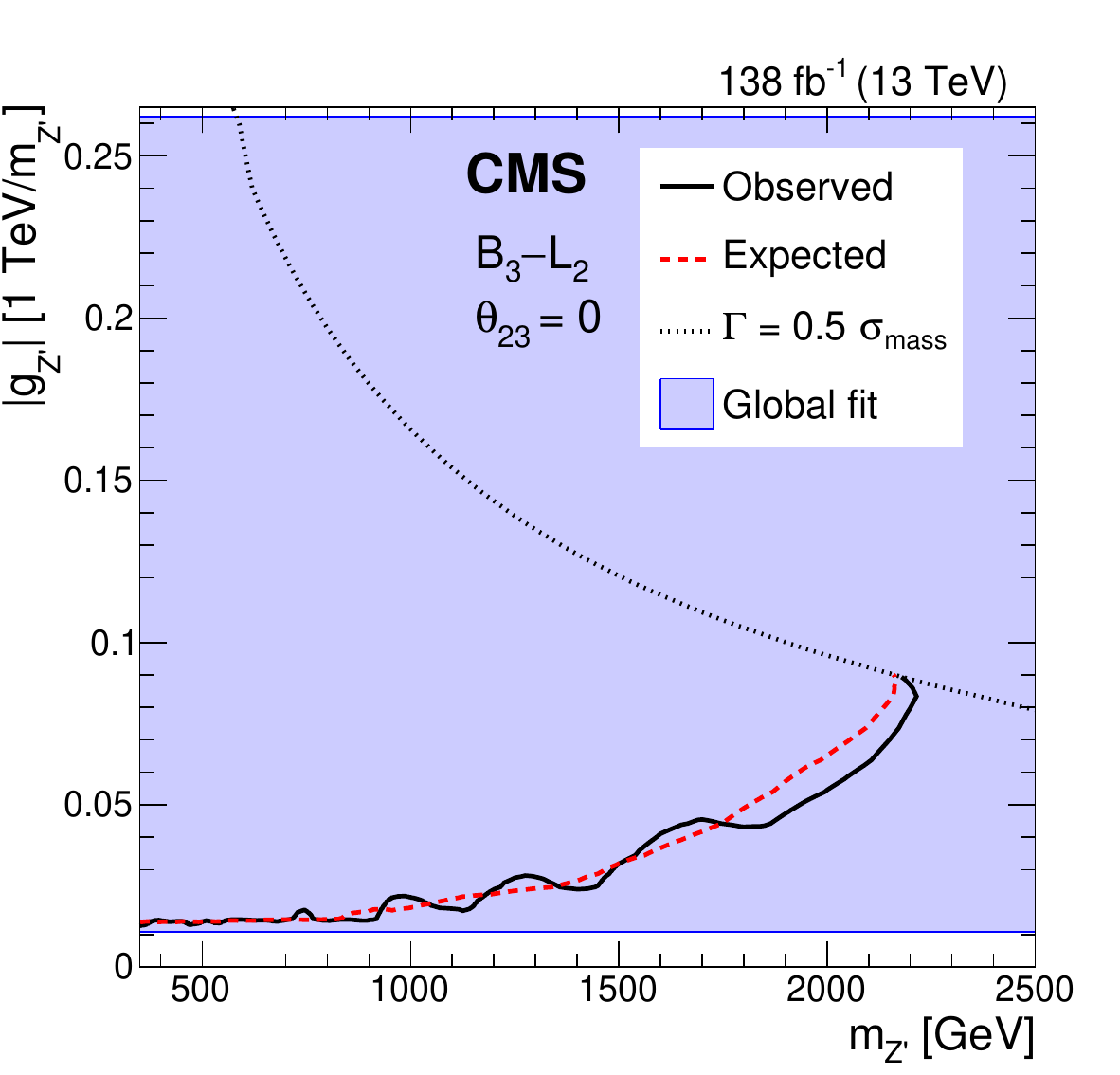}
  \caption{Exclusion limits at 95\%~\CL in the $\abs{\gZp}$--$\mZp$
    plane for the $B_{3}\!-\!L_{2}$ model~\cite{Allanach:2022iod}
    for a fixed value of $\ttwothree=0$.
    The solid black (dashed red) curve represents the observed (median expected) exclusion.
    The dotted curve denotes the coupling values at which the \Zprime width
    equals one half of the $\Pgm\Pgm$ invariant mass resolution.
    For larger values of the couplings, the narrow width approximation intrinsic to the search strategy is not considered valid.
    The region enclosed between the solid black (dashed red) and the dotted curves is 
    (expected to be) excluded.
    The shaded blue area represents the $\abs{\gZp}$ range preferred from the global fit in Ref.~\cite{Allanach:2022iod} at 95\%~\CL.
  }
  \label{fig:limits_b3ml2_t23eq0}
\end{figure}

The results are also used to set constraints on the $B_{3}\!-\!L_{2}$ model from Ref.~\cite{Allanach:2022iod}.
As in the case of the LFU model, we restrict our exclusion ranges 
to regions with
$\Gamma(\Zprime) < \sigma_\text{mass} / 2$.
As shown in Fig.~\ref{fig:limits_b3ml2},
most of the allowed parameter space is excluded for a $\Zprime$ boson with $350<\mZp<500\GeV$,
while the constraints are less stringent for higher $\mZp$.
Since exclusion limits are nearly independent of the \ttwothree parameter in the $B_{3}\!-\!L_{2}$ model
or, equivalently, $\delta_{\PQb\PQs}$ in the LFU model,
we also show in Fig.~\ref{fig:limits_b3ml2_t23eq0} exclusion regions
in the $\abs{\gZp}$--$\mZp$ plane for a fixed value of $\ttwothree=0$.

\clearpage

\section{Summary}
\label{sec:Summary}

A search for high-mass dimuon resonance production in association with one or more $\PQb$ quark jets has been presented,
using data collected with the CMS experiment at the LHC
that correspond to an integrated luminosity of \lumiT at a center-of-mass energy of $13\TeV$.

The limits are derived on the total number of signal events
with $\Nb=1$ and $\geq$2,
where \Nb denotes the multiplicity of $\PQb$ quark jets,
and these limits are model independent to the extent that the signal arises from the direct production and subsequent decay of a narrow dimuon resonance.
The relative fraction of events with $\Nb\geq2$ is varied
to probe a range of hypotheses of signal production in association with $\PQb$ quarks.
The limits are presented as a function of the analyzed dimuon resonance mass values.

Results are also interpreted in terms of a lepton-flavor-universal model that involves $\Zprime$ boson couplings
to \PQb quarks ($\gb$) and muons,
where the $\Zprime$ boson couplings to all neutrinos ($g_{\PGn}$) and to all charged leptons ($\gl$) are assumed to be equal, 
the $\gb$ coupling scales both $\Zprime\PQb\PQb$ and $\Zprime\PQs\PQb$ interactions,
and the separate $\delta_{\PQb\PQs}$ coupling solely scales the $\Zprime\PQs\PQb$ interaction.
The exclusions in this model are presented in terms of the coupling strengths $\gl$ and $\gb$, and the mass of the \Zprime boson ($\mZp$).
For a $\Zprime$ boson with $\mZp=350\GeV$ (2\TeV) and $\abs{\delta_{\PQb\PQs}} < 0.25$,
the majority of the parameter space with $0.0057 < \abs{\gl} < 0.35$ ($0.25 < \abs{\gl} < 0.43$)
and $0.0079 < \abs{g_{\PQb}} < 0.46$ ($0.34 < \abs{g_{\PQb}} < 0.57$) is excluded
at $95\%$ confidence level.

Constraints are also set on
a specific $\Zprime$ model ({$B_3\!-\!L_2$}), constructed to accommodate
possible contributions to $\PQb \to \PQs \ell^{-} \ell^{+}$ transitions beyond the standard model.
In this scenario, most of the allowed parameter space is excluded for a $\Zprime$ boson with $350<\mZp<500\GeV$,
while the constraints are less stringent for higher $\mZp$ hypotheses.

This is the first dedicated search at the LHC for a high-mass, narrow dimuon resonance produced in association
with multiple $\PQb$ quark jets, and the constraints obtained on models with this signature are the most stringent to date.

\begin{acknowledgments}
We congratulate our colleagues in the CERN accelerator departments for the excellent performance of the LHC and thank the technical and administrative staffs at CERN and at other CMS institutes for their contributions to the success of the CMS effort. In addition, we gratefully acknowledge the computing centers and personnel of the Worldwide LHC Computing Grid and other centers for delivering so effectively the computing infrastructure essential to our analyses. Finally, we acknowledge the enduring support for the construction and operation of the LHC, the CMS detector, and the supporting computing infrastructure provided by the following funding agencies: SC (Armenia), BMBWF and FWF (Austria); FNRS and FWO (Belgium); CNPq, CAPES, FAPERJ, FAPERGS, and FAPESP (Brazil); MES and BNSF (Bulgaria); CERN; CAS, MoST, and NSFC (China); MINCIENCIAS (Colombia); MSES and CSF (Croatia); RIF (Cyprus); SENESCYT (Ecuador); MoER, ERC PUT and ERDF (Estonia); Academy of Finland, MEC, and HIP (Finland); CEA and CNRS/IN2P3 (France); BMBF, DFG, and HGF (Germany); GSRI (Greece); NKFIH (Hungary); DAE and DST (India); IPM (Iran); SFI (Ireland); INFN (Italy); MSIP and NRF (Republic of Korea); MES (Latvia); LAS (Lithuania); MOE and UM (Malaysia); BUAP, CINVESTAV, CONACYT, LNS, SEP, and UASLP-FAI (Mexico); MOS (Montenegro); MBIE (New Zealand); PAEC (Pakistan); MES and NSC (Poland); FCT (Portugal);MESTD (Serbia); MCIN/AEI and PCTI (Spain); MOSTR (Sri Lanka); Swiss Funding Agencies (Switzerland); MST (Taipei); MHESI and NSTDA (Thailand); TUBITAK and TENMAK (Turkey); NASU (Ukraine); STFC (United Kingdom); DOE and NSF (USA).

\hyphenation{Rachada-pisek} Individuals have received support from the Marie-Curie program and the European Research Council and Horizon 2020 Grant, contract Nos.\ 675440, 724704, 752730, 758316, 765710, 824093, 884104, and COST Action CA16108 (European Union); the Leventis Foundation; the Alfred P.\ Sloan Foundation; the Alexander von Humboldt Foundation; the Science Committee, project no. 22rl-037 (Armenia); the Belgian Federal Science Policy Office; the Fonds pour la Formation \`a la Recherche dans l'Industrie et dans l'Agriculture (FRIA-Belgium); the Agentschap voor Innovatie door Wetenschap en Technologie (IWT-Belgium); the F.R.S.-FNRS and FWO (Belgium) under the ``Excellence of Science -- EOS" -- be.h project n.\ 30820817; the Beijing Municipal Science \& Technology Commission, No. Z191100007219010; the Ministry of Education, Youth and Sports (MEYS) of the Czech Republic; the Hellenic Foundation for Research and Innovation (HFRI), Project Number 2288 (Greece); the Deutsche Forschungsgemeinschaft (DFG), under Germany's Excellence Strategy -- EXC 2121 ``Quantum Universe" -- 390833306, and under project number 400140256 - GRK2497; the Hungarian Academy of Sciences, the New National Excellence Program - \'UNKP, the NKFIH research grants K 124845, K 124850, K 128713, K 128786, K 129058, K 131991, K 133046, K 138136, K 143460, K 143477, 2020-2.2.1-ED-2021-00181, and TKP2021-NKTA-64 (Hungary); the Council of Science and Industrial Research, India; the Latvian Council of Science; the Ministry of Education and Science, project no. 2022/WK/14, and the National Science Center, contracts Opus 2021/41/B/ST2/01369 and 2021/43/B/ST2/01552 (Poland); the Funda\c{c}\~ao para a Ci\^encia e a Tecnologia, grant CEECIND/01334/2018 (Portugal); the National Priorities Research Program by Qatar National Research Fund; MCIN/AEI/10.13039/501100011033, ERDF ``a way of making Europe", and the Programa Estatal de Fomento de la Investigaci{\'o}n Cient{\'i}fica y T{\'e}cnica de Excelencia Mar\'{\i}a de Maeztu, grant MDM-2017-0765 and Programa Severo Ochoa del Principado de Asturias (Spain); the Chulalongkorn Academic into Its 2nd Century Project Advancement Project, and the National Science, Research and Innovation Fund via the Program Management Unit for Human Resources \& Institutional Development, Research and Innovation, grant B05F650021 (Thailand); the Kavli Foundation; the Nvidia Corporation; the SuperMicro Corporation; the Welch Foundation, contract C-1845; and the Weston Havens Foundation (USA).
\end{acknowledgments}

\bibliography{auto_generated}
\cleardoublepage \appendix\section{The CMS Collaboration \label{app:collab}}\begin{sloppypar}\hyphenpenalty=5000\widowpenalty=500\clubpenalty=5000\input{EXO-22-016-public-authorlist.tex}\end{sloppypar}
\end{document}

%% file: EXO-22-016-public-authorlist.tex
\cmsinstitute{Yerevan Physics Institute, Yerevan, Armenia}
{\tolerance=6000
A.~Hayrapetyan, A.~Tumasyan\cmsAuthorMark{1}\cmsorcid{0009-0000-0684-6742}
\par}
\cmsinstitute{Institut f\"{u}r Hochenergiephysik, Vienna, Austria}
{\tolerance=6000
W.~Adam\cmsorcid{0000-0001-9099-4341}, J.W.~Andrejkovic, T.~Bergauer\cmsorcid{0000-0002-5786-0293}, S.~Chatterjee\cmsorcid{0000-0003-2660-0349}, K.~Damanakis\cmsorcid{0000-0001-5389-2872}, M.~Dragicevic\cmsorcid{0000-0003-1967-6783}, A.~Escalante~Del~Valle\cmsorcid{0000-0002-9702-6359}, P.S.~Hussain\cmsorcid{0000-0002-4825-5278}, M.~Jeitler\cmsAuthorMark{2}\cmsorcid{0000-0002-5141-9560}, N.~Krammer\cmsorcid{0000-0002-0548-0985}, D.~Liko\cmsorcid{0000-0002-3380-473X}, I.~Mikulec\cmsorcid{0000-0003-0385-2746}, J.~Schieck\cmsAuthorMark{2}\cmsorcid{0000-0002-1058-8093}, R.~Sch\"{o}fbeck\cmsorcid{0000-0002-2332-8784}, D.~Schwarz\cmsorcid{0000-0002-3821-7331}, M.~Sonawane\cmsorcid{0000-0003-0510-7010}, S.~Templ\cmsorcid{0000-0003-3137-5692}, W.~Waltenberger\cmsorcid{0000-0002-6215-7228}, C.-E.~Wulz\cmsAuthorMark{2}\cmsorcid{0000-0001-9226-5812}
% ...
\par}
\vskip\cmsinstskip
\dag:~Deceased\\
$^{1}$Also at Yerevan State University, Yerevan, Armenia\\
$^{2}$Also at TU Wien, Vienna, Austria\\
% ...


%% file: EXO-22-016_temp.bbl
\providecommand{\href}[2]{#2}\begingroup\raggedright\begin{thebibliography}{10}%
\makeatletter
\providecommand{\hrefCMSnoop }[0]{\@secondoftwo}%
\makeatother
\providecommand{\doi}{\texttt{doi:}\begingroup \urlstyle{tt}\Url}

\bibitem{LHCb:2013ghj}
\hrefCMSnoop {}{{LHCb Collaboration}, ``{Measurement of form-factor-independent
  observables in the decay $\PBz \to \HepParticle{\PK}{}{\ast 0} \PGmp
  \PGmm$}'',} \textit{ Phys. Rev. Lett.} \textbf{ 111} (2013) 191801,
  \href{http://dx.doi.org/10.1103/PhysRevLett.111.191801}{\doi{10.1103/PhysRevLett.111.191801}},
  \href{http://www.arXiv.org/abs/1308.1707}{\texttt{arXiv:1308.1707}}.

\bibitem{LHCb:2014cxe}
\hrefCMSnoop {}{{LHCb Collaboration}, ``{Differential branching fractions and
  isospin asymmetries of $\PB \to {\HepParticle{\PK}{}{(\ast)}} \PGmp\PGmm$
  decays}'',} \textit{ JHEP} \textbf{ 06} (2014) 133,
  \href{http://dx.doi.org/10.1007/JHEP06(2014)133}{\doi{10.1007/JHEP06(2014)133}},
  \href{http://www.arXiv.org/abs/1403.8044}{\texttt{arXiv:1403.8044}}.

\bibitem{LHCb:2015wdu}
\hrefCMSnoop {}{{LHCb Collaboration}, ``{Angular analysis and differential
  branching fraction of the decay $\PBzs\to\PGf\PGmp\PGmm$}'',} \textit{ JHEP}
  \textbf{ 09} (2015) 179,
  \href{http://dx.doi.org/10.1007/JHEP09(2015)179}{\doi{10.1007/JHEP09(2015)179}},
  \href{http://www.arXiv.org/abs/1506.08777}{\texttt{arXiv:1506.08777}}.

\bibitem{LHCb:2015svh}
\hrefCMSnoop {}{{LHCb Collaboration}, ``{Angular analysis of the $\PBz \to
  \HepParticle{\PK}{}{\ast 0} \PGmp \PGmm$ decay using $3\fbinv$ of integrated
  luminosity}'',} \textit{ JHEP} \textbf{ 02} (2016) 104,
  \href{http://dx.doi.org/10.1007/JHEP02(2016)104}{\doi{10.1007/JHEP02(2016)104}},
  \href{http://www.arXiv.org/abs/1512.04442}{\texttt{arXiv:1512.04442}}.

\bibitem{CMS:2015bcy}
\hrefCMSnoop {}{{CMS Collaboration}, ``{Angular analysis of the decay $\PBz \to
  \HepParticle{\PK}{}{\ast 0} \PGmp \PGmm$ from $\Pp\Pp$ collisions at
  $\sqrt{s} = 8\TeV$}'',} \textit{ Phys. Lett. B} \textbf{ 753} (2016) 424,
  \href{http://dx.doi.org/10.1016/j.physletb.2015.12.020}{\doi{10.1016/j.physletb.2015.12.020}},
  \href{http://www.arXiv.org/abs/1507.08126}{\texttt{arXiv:1507.08126}}.

\bibitem{LHCb:2017avl}
\hrefCMSnoop {}{{LHCb Collaboration}, ``{Test of lepton universality with $\PBz
  \rightarrow {\HepParticle{\PK}{}{\ast 0}} \ell^{+}\ell^{-}$ decays}'',}
  \textit{ JHEP} \textbf{ 08} (2017) 055,
  \href{http://dx.doi.org/10.1007/JHEP08(2017)055}{\doi{10.1007/JHEP08(2017)055}},
  \href{http://www.arXiv.org/abs/1705.05802}{\texttt{arXiv:1705.05802}}.

\bibitem{Bobeth:2017vxj}
\hrefCMSnoop {}{C.~Bobeth, M.~Chrzaszcz, D.~van Dyk, and J.~Virto,
  ``{Long-distance effects in $\PB\to \HepParticle{\PK}{}{\ast}\ell \ell$ from
  analyticity}'',} \textit{ Eur. Phys. J. C} \textbf{ 78} (2018) 451,
  \href{http://dx.doi.org/10.1140/epjc/s10052-018-5918-6}{\doi{10.1140/epjc/s10052-018-5918-6}},
  \href{http://www.arXiv.org/abs/1707.07305}{\texttt{arXiv:1707.07305}}.

\bibitem{CMS:2017rzx}
\hrefCMSnoop {}{{CMS Collaboration}, ``{Measurement of angular parameters from
  the decay $\PBz \to \HepParticle{\PK}{}{\ast 0} \PGmp \PGmm$ in proton-proton
  collisions at $\sqrt{s} = 8\TeV$}'',} \textit{ Phys. Lett. B} \textbf{ 781}
  (2018) 517,
  \href{http://dx.doi.org/10.1016/j.physletb.2018.04.030}{\doi{10.1016/j.physletb.2018.04.030}},
  \href{http://www.arXiv.org/abs/1710.02846}{\texttt{arXiv:1710.02846}}.

\bibitem{ATLAS:2018gqc}
\hrefCMSnoop {}{{ATLAS Collaboration}, ``{Angular analysis of $\PBzd \to
  \HepParticle{\PK}{}{\ast}\PGmp\PGmm$ decays in $\Pp\Pp$ collisions at
  $\sqrt{s}= 8\TeV$ with the ATLAS detector}'',} \textit{ JHEP} \textbf{ 10}
  (2018) 047,
  \href{http://dx.doi.org/10.1007/JHEP10(2018)047}{\doi{10.1007/JHEP10(2018)047}},
  \href{http://www.arXiv.org/abs/1805.04000}{\texttt{arXiv:1805.04000}}.

\bibitem{LHCb:2019hip}
\hrefCMSnoop {}{{LHCb Collaboration}, ``{Search for lepton-universality
  violation in $\PBp \to \PKp \ell^+\ell^-$ decays}'',} \textit{ Phys. Rev.
  Lett.} \textbf{ 122} (2019) 191801,
  \href{http://dx.doi.org/10.1103/PhysRevLett.122.191801}{\doi{10.1103/PhysRevLett.122.191801}},
  \href{http://www.arXiv.org/abs/1903.09252}{\texttt{arXiv:1903.09252}}.

\bibitem{LHCb:2021trn}
\hrefCMSnoop {}{{LHCb Collaboration}, ``{Test of lepton universality in
  beauty-quark decays}'',} \textit{ Nature Phys.} \textbf{ 18} (2022) 277,
  \href{http://dx.doi.org/10.1038/s41567-021-01478-8}{\doi{10.1038/s41567-021-01478-8}},
  \href{http://www.arXiv.org/abs/2103.11769}{\texttt{arXiv:2103.11769}}.

\bibitem{LHCb:2021lvy}
\hrefCMSnoop {}{{LHCb Collaboration}, ``{Tests of lepton universality using
  $\PBz \to \PKzS \ell^+ \ell^-$ and $\PBp \to {\HepParticle{\PK}{}{\ast +}}
  \ell^+ \ell^-$ decays}'',} \textit{ Phys. Rev. Lett.} \textbf{ 128} (2022)
  191802,
  \href{http://dx.doi.org/10.1103/PhysRevLett.128.191802}{\doi{10.1103/PhysRevLett.128.191802}},
  \href{http://www.arXiv.org/abs/2110.09501}{\texttt{arXiv:2110.09501}}.

\bibitem{Parrott:2022rgu}
\hrefCMSnoop {}{{HPQCD} Collaboration, ``{$\PB\to\PK$ and $\PD\to\PK$ form
  factors from fully relativistic lattice QCD}'',} \textit{ Phys. Rev. D}
  \textbf{ 107} (2023) 014510,
  \href{http://dx.doi.org/10.1103/PhysRevD.107.014510}{\doi{10.1103/PhysRevD.107.014510}},
  \href{http://www.arXiv.org/abs/2207.12468}{\texttt{arXiv:2207.12468}}.

\bibitem{Leike:1998wr}
\hrefCMSnoop {}{A.~Leike, ``{The phenomenology of extra neutral gauge
  bosons}'',} \textit{ Phys. Rept.} \textbf{ 317} (1999) 143,
  \href{http://dx.doi.org/10.1016/S0370-1573(98)00133-1}{\doi{10.1016/S0370-1573(98)00133-1}},
  \href{http://www.arXiv.org/abs/hep-ph/9805494}{\texttt{arXiv:hep-ph/9805494}}.

\bibitem{Langacker:2008yv}
\hrefCMSnoop {}{P.~Langacker, ``{The physics of heavy $\Zprime$ gauge
  bosons}'',} \textit{ Rev. Mod. Phys.} \textbf{ 81} (2009) 1199,
  \href{http://dx.doi.org/10.1103/RevModPhys.81.1199}{\doi{10.1103/RevModPhys.81.1199}},
  \href{http://www.arXiv.org/abs/0801.1345}{\texttt{arXiv:0801.1345}}.

\bibitem{LHCb:2022qnv}
\hrefCMSnoop {}{{LHCb Collaboration}, ``{Test of lepton universality in $\PQb
  \rightarrow \PQs \ell^+ \ell^-$ decays}'',} \textit{ Phys. Rev. Lett.}
  \textbf{ 131} (2023) 051803,
  \href{http://dx.doi.org/10.1103/PhysRevLett.131.051803}{\doi{10.1103/PhysRevLett.131.051803}},
  \href{http://www.arXiv.org/abs/2212.09152}{\texttt{arXiv:2212.09152}}.

\bibitem{LHCb:2022zom}
\hrefCMSnoop {}{{LHCb Collaboration}, ``{Measurement of lepton universality
  parameters in $\PBp \to \PKp \ell^+\ell^-$ and $\PBz \to
  {\HepParticle{\PK}{}{\ast 0}} \ell^+\ell^-$ decays}'',} \textit{ Phys. Rev.
  D} \textbf{ 108} (2023) 032002,
  \href{http://dx.doi.org/10.1103/PhysRevD.108.032002}{\doi{10.1103/PhysRevD.108.032002}},
  \href{http://www.arXiv.org/abs/2212.09153}{\texttt{arXiv:2212.09153}}.

\bibitem{Allanach:2022iod}
\hrefCMSnoop {}{B.~Allanach and J.~Davighi, ``{The Rumble in the Meson: a
  leptoquark versus a $\Zprime$ to fit $\PQb \to \PQs \PGmp \PGmm$ anomalies
  including 2022 LHCb $\textrm{R}_{\HepParticle{\PK}{}{(\ast)}}$
  measurements}'',} \textit{ JHEP} \textbf{ 04} (2023) 033,
  \href{http://dx.doi.org/10.1007/JHEP04(2023)033}{\doi{10.1007/JHEP04(2023)033}},
  \href{http://www.arXiv.org/abs/2211.11766}{\texttt{arXiv:2211.11766}}.

\bibitem{ATLAS:2019erb}
\hrefCMSnoop {}{{ATLAS Collaboration}, ``{Search for high-mass dilepton
  resonances using $139\fbinv$ of $\Pp{}\Pp{}$ collision data collected at
  $\sqrt{s}=13\TeV$ with the ATLAS detector}'',} \textit{ Phys. Lett. B}
  \textbf{ 796} (2019) 68,
  \href{http://dx.doi.org/10.1016/j.physletb.2019.07.016}{\doi{10.1016/j.physletb.2019.07.016}},
  \href{http://www.arXiv.org/abs/1903.06248}{\texttt{arXiv:1903.06248}}.

\bibitem{EXO-19-019}
\hrefCMSnoop {}{{CMS Collaboration}, ``{Search for resonant and nonresonant new
  phenomena in high-mass dilepton final states at $\sqrt{s}=13\TeV$}'',}
  \textit{ JHEP} \textbf{ 07} (2021) 208,
  \href{http://dx.doi.org/10.1007/JHEP07(2021)208}{\doi{10.1007/JHEP07(2021)208}},
  \href{http://www.arXiv.org/abs/2103.02708}{\texttt{arXiv:2103.02708}}.

\bibitem{ParticleDataGroup:2022pth}
\hrefCMSnoop {}{{Particle Data Group}, R.~L. Workman {et~al.}, ``{Review of
  Particle Physics}'',} \textit{ PTEP} \textbf{ 2022} (2022) 083C01,
  \href{http://dx.doi.org/10.1093/ptep/ptac097}{\doi{10.1093/ptep/ptac097}}.

\bibitem{Alonso:2017uky}
\hrefCMSnoop {}{R.~Alonso, P.~Cox, C.~Han, and T.~T. Yanagida, ``{Flavoured
  {$B\!-\!L$} local symmetry and anomalous rare $\PB$ decays}'',} \textit{
  Phys. Lett. B} \textbf{ 774} (2017) 643,
  \href{http://dx.doi.org/10.1016/j.physletb.2017.10.027}{\doi{10.1016/j.physletb.2017.10.027}},
  \href{http://www.arXiv.org/abs/1705.03858}{\texttt{arXiv:1705.03858}}.

\bibitem{Bonilla:2017lsq}
\hrefCMSnoop {}{C.~Bonilla, T.~Modak, R.~Srivastava, and J.~W.~F. Valle,
  ``{$U(1)_{B_3-3L_\PGm}$ gauge symmetry as a simple description of
  $\PQb\to\PQs$ anomalies}'',} \textit{ Phys. Rev. D} \textbf{ 98} (2018)
  095002,
  \href{http://dx.doi.org/10.1103/PhysRevD.98.095002}{\doi{10.1103/PhysRevD.98.095002}},
  \href{http://www.arXiv.org/abs/1705.00915}{\texttt{arXiv:1705.00915}}.

\bibitem{ATLAS:2019odt}
\hrefCMSnoop {}{{ATLAS Collaboration}, ``{Search for scalar resonances decaying
  into $\PGmp{}\PGmm{}$ in events with and without $\PQb$-tagged jets produced
  in proton-proton collisions at $\sqrt{s}=13\TeV$ with the ATLAS detector}'',}
  \textit{ JHEP} \textbf{ 07} (2019) 117,
  \href{http://dx.doi.org/10.1007/JHEP07(2019)117}{\doi{10.1007/JHEP07(2019)117}},
  \href{http://www.arXiv.org/abs/1901.08144}{\texttt{arXiv:1901.08144}}.

\bibitem{CMS:2019mij}
\hrefCMSnoop {}{{CMS Collaboration}, ``{Search for MSSM Higgs bosons decaying
  to $\PGmp{}\PGmm{}$ in proton-proton collisions at $\sqrt{s}=13\TeV$}'',}
  \textit{ Phys. Lett. B} \textbf{ 798} (2019) 134992,
  \href{http://dx.doi.org/10.1016/j.physletb.2019.134992}{\doi{10.1016/j.physletb.2019.134992}},
  \href{http://www.arXiv.org/abs/1907.03152}{\texttt{arXiv:1907.03152}}.

\bibitem{ATLAS:2021mla}
\hrefCMSnoop {}{{ATLAS Collaboration}, ``{Search for new phenomena in final
  states with two leptons and one or no $\PQb$-tagged jets at $\sqrt{s} =
  13\TeV$ using the ATLAS detector}'',} \textit{ Phys. Rev. Lett.} \textbf{
  127} (2021) 141801,
  \href{http://dx.doi.org/10.1103/PhysRevLett.127.141801}{\doi{10.1103/PhysRevLett.127.141801}},
  \href{http://www.arXiv.org/abs/2105.13847}{\texttt{arXiv:2105.13847}}.

\bibitem{CMS:2021sch}
\hrefCMSnoop {}{{CMS Collaboration}, ``{Search for long-lived particles
  decaying into muon pairs in proton-proton collisions at $\sqrt{s}=13\TeV$
  collected with a dedicated high-rate data stream}'',} \textit{ JHEP} \textbf{
  04} (2022) 062,
  \href{http://dx.doi.org/10.1007/JHEP04(2022)062}{\doi{10.1007/JHEP04(2022)062}},
  \href{http://www.arXiv.org/abs/2112.13769}{\texttt{arXiv:2112.13769}}.

\bibitem{hepdata}
\hrefCMSnoop {}{}{HEPD}ata record for this analysis, 2023.
\newblock
  \href{http://dx.doi.org/10.17182/hepdata.141455}{\doi{10.17182/hepdata.141455}}.

\bibitem{Chatrchyan:2008zzk}
\hrefCMSnoop {}{{CMS Collaboration}, ``The {CMS} experiment at the {CERN}
  {LHC}'',} \textit{ JINST} \textbf{ 3} (2008) S08004,
\href{http://dx.doi.org/10.1088/1748-0221/3/08/S08004}{\doi{10.1088/1748-0221/3/08/S08004}}.

\bibitem{CMS:2020cmk}
\hrefCMSnoop {}{{CMS Collaboration}, ``{Performance of the CMS Level-1 trigger
  in proton-proton collisions at $\sqrt{s} = 13$\,TeV}'',} \textit{ JINST}
  \textbf{ 15} (2020) P10017,
  \href{http://dx.doi.org/10.1088/1748-0221/15/10/P10017}{\doi{10.1088/1748-0221/15/10/P10017}},
  \href{http://www.arXiv.org/abs/2006.10165}{\texttt{arXiv:2006.10165}}.

\bibitem{CMS:2016ngn}
\hrefCMSnoop {}{{CMS Collaboration}, ``{The CMS trigger system}'',} \textit{
  JINST} \textbf{ 12} (2017) P01020,
  \href{http://dx.doi.org/10.1088/1748-0221/12/01/P01020}{\doi{10.1088/1748-0221/12/01/P01020}},
\href{http://www.arXiv.org/abs/1609.02366}{\texttt{arXiv:1609.02366}}.

\bibitem{Sirunyan:2017ulk}
\hrefCMSnoop {}{{CMS Collaboration}, ``{Particle-flow reconstruction and global
  event description with the CMS detector}'',} \textit{ JINST} \textbf{ 12}
  (2017) P10003,
  \href{http://dx.doi.org/10.1088/1748-0221/12/10/P10003}{\doi{10.1088/1748-0221/12/10/P10003}},
\href{http://www.arXiv.org/abs/1706.04965}{\texttt{arXiv:1706.04965}}.

\bibitem{CMS-TDR-15-02}
\href {http://cds.cern.ch/record/2020886}{{CMS Collaboration}, ``Technical
  proposal for the {Phase-II} upgrade of the {Compact Muon Solenoid}'',} CMS
  Technical Proposal CERN-LHCC-2015-010, CMS-TDR-15-02, 2015.

\bibitem{Sirunyan:2018fpa}
\hrefCMSnoop {}{{CMS Collaboration}, ``{Performance of the CMS muon detector
  and muon reconstruction with proton-proton collisions at
  $\sqrt{s}=13\TeV$}'',} \textit{ JINST} \textbf{ 13} (2018) P06015,
  \href{http://dx.doi.org/10.1088/1748-0221/13/06/P06015}{\doi{10.1088/1748-0221/13/06/P06015}},
  \href{http://www.arXiv.org/abs/1804.04528}{\texttt{arXiv:1804.04528}}.

\bibitem{CMS:2019ied}
\hrefCMSnoop {}{{CMS Collaboration}, ``{Performance of the reconstruction and
  identification of high-momentum muons in proton-proton collisions at
  $\sqrt{s} =13\TeV$}'',} \textit{ JINST} \textbf{ 15} (2020) P02027,
  \href{http://dx.doi.org/10.1088/1748-0221/15/02/P02027}{\doi{10.1088/1748-0221/15/02/P02027}},
  \href{http://www.arXiv.org/abs/1912.03516}{\texttt{arXiv:1912.03516}}.

\bibitem{CMS:2010svw}
\hrefCMSnoop {}{{CMS Collaboration}, ``{Measurements of inclusive $\PW$ and
  $\PZ$ cross sections in $\Pp{}\Pp{}$ collisions at $\sqrt{s}=7\TeV$}'',}
  \textit{ JHEP} \textbf{ 01} (2011) 080,
  \href{http://dx.doi.org/10.1007/JHEP01(2011)080}{\doi{10.1007/JHEP01(2011)080}},
  \href{http://www.arXiv.org/abs/1012.2466}{\texttt{arXiv:1012.2466}}.

\bibitem{CMS:2021yvr}
\hrefCMSnoop {}{{CMS Collaboration}, ``{Performance of the CMS muon trigger
  system in proton-proton collisions at $\sqrt{s} = 13\TeV$}'',} \textit{
  JINST} \textbf{ 16} (2021) P07001,
  \href{http://dx.doi.org/10.1088/1748-0221/16/07/P07001}{\doi{10.1088/1748-0221/16/07/P07001}},
  \href{http://www.arXiv.org/abs/2102.04790}{\texttt{arXiv:2102.04790}}.

\bibitem{CMS:2020uim}
\hrefCMSnoop {}{{CMS Collaboration}, ``{Electron and photon reconstruction and
  identification with the CMS experiment at the CERN LHC}'',} \textit{ JINST}
  \textbf{ 16} (2021) P05014,
  \href{http://dx.doi.org/10.1088/1748-0221/16/05/P05014}{\doi{10.1088/1748-0221/16/05/P05014}},
  \href{http://www.arXiv.org/abs/2012.06888}{\texttt{arXiv:2012.06888}}.

\bibitem{Cacciari:2008gp}
\hrefCMSnoop {}{M.~Cacciari, G.~P. Salam, and G.~Soyez, ``{The anti-$\kt$ jet
  clustering algorithm}'',} \textit{ JHEP} \textbf{ 04} (2008) 063,
  \href{http://dx.doi.org/10.1088/1126-6708/2008/04/063}{\doi{10.1088/1126-6708/2008/04/063}},
  \href{http://www.arXiv.org/abs/0802.1189}{\texttt{arXiv:0802.1189}}.

\bibitem{Cacciari:2011ma}
\hrefCMSnoop {}{M.~Cacciari, G.~P. Salam, and G.~Soyez, ``{FastJet user
  manual}'',} \textit{ Eur. Phys. J. C} \textbf{ 72} (2012) 1896,
  \href{http://dx.doi.org/10.1140/epjc/s10052-012-1896-2}{\doi{10.1140/epjc/s10052-012-1896-2}},
\href{http://www.arXiv.org/abs/1111.6097}{\texttt{arXiv:1111.6097}}.

\bibitem{Sirunyan:2020foa}
\hrefCMSnoop {}{{CMS Collaboration}, ``{Pileup mitigation at CMS in 13 TeV
  data}'',} \textit{ JINST} \textbf{ 15} (2020) P09018,
  \href{http://dx.doi.org/10.1088/1748-0221/15/09/p09018}{\doi{10.1088/1748-0221/15/09/p09018}},
  \href{http://www.arXiv.org/abs/2003.00503}{\texttt{arXiv:2003.00503}}.

\bibitem{Bols:2020bkb}
E.~Bols\hrefCMSnoop {}{ {et~al.}, ``{Jet flavour classification using
  {DeepJet}}'',} \textit{ JINST} \textbf{ 15} (2020) 12012,
  \href{http://dx.doi.org/10.1088/1748-0221/15/12/P12012}{\doi{10.1088/1748-0221/15/12/P12012}},
  \href{http://www.arXiv.org/abs/2008.10519}{\texttt{arXiv:2008.10519}}.

\bibitem{CMS-DP-2023-005}
\href {https://cds.cern.ch/record/2854609}{{CMS Collaboration}, ``{Performance
  summary of AK4 jet $\PQb$ tagging with data from proton-proton collisions at
  $13\TeV$ with the CMS detector}'',} {CMS Detector Performance Note}
  {CMS-DP-2023-005}, 2023.

\bibitem{Bertolini:2014bba}
\hrefCMSnoop {}{D.~Bertolini, P.~Harris, M.~Low, and N.~Tran, ``Pileup per
  particle identification'',} \textit{ JHEP} \textbf{ 10} (2014) 059,
  \href{http://dx.doi.org/10.1007/JHEP10(2014)059}{\doi{10.1007/JHEP10(2014)059}},
\href{http://www.arXiv.org/abs/1407.6013}{\texttt{arXiv:1407.6013}}.

\bibitem{CMS:2019ctu}
\hrefCMSnoop {}{{CMS Collaboration}, ``{Performance of missing transverse
  momentum reconstruction in proton-proton collisions at $\sqrt{s} = 13\TeV$
  using the {CMS} detector}'',} \textit{ JINST} \textbf{ 14} (2019) P07004,
  \href{http://dx.doi.org/10.1088/1748-0221/14/07/P07004}{\doi{10.1088/1748-0221/14/07/P07004}},
\href{http://www.arXiv.org/abs/1903.06078}{\texttt{arXiv:1903.06078}}.

\bibitem{Frixione:2007vw}
\hrefCMSnoop {}{S.~Frixione, P.~Nason, and C.~Oleari, ``{Matching NLO QCD
  computations with parton shower simulations: the POWHEG method}'',} \textit{
  JHEP} \textbf{ 11} (2007) 070,
  \href{http://dx.doi.org/10.1088/1126-6708/2007/11/070}{\doi{10.1088/1126-6708/2007/11/070}},
\href{http://www.arXiv.org/abs/0709.2092}{\texttt{arXiv:0709.2092}}.

\bibitem{Alioli:2009je}
\hrefCMSnoop {}{S.~Alioli, P.~Nason, C.~Oleari, and E.~Re, ``{NLO single-top
  production matched with shower in POWHEG: $s$- and $t$-channel
  contributions}'',} \textit{ JHEP} \textbf{ 09} (2009) 111,
  \href{http://dx.doi.org/10.1088/1126-6708/2009/09/111}{\doi{10.1088/1126-6708/2009/09/111}},
  \href{http://www.arXiv.org/abs/0907.4076}{\texttt{arXiv:0907.4076}}.
[Erratum: \DOI{10.1007/JHEP02(2010)011}].

\bibitem{Melia:2011tj}
\hrefCMSnoop {}{T.~Melia, P.~Nason, R.~R{\"o}ntsch, and G.~Zanderighi,
  ``{$\PWp\PWm$ , $\PW\PZ$ and $\PZ\PZ$ production in the POWHEG BOX}'',}
  \textit{ JHEP} \textbf{ 11} (2011) 078,
  \href{http://dx.doi.org/10.1007/JHEP11(2011)078}{\doi{10.1007/JHEP11(2011)078}},
  \href{http://www.arXiv.org/abs/1107.5051}{\texttt{arXiv:1107.5051}}.

\bibitem{Nason:2013ydw}
\hrefCMSnoop {}{P.~Nason and G.~Zanderighi, ``{$\PWp\PWm$, $\PW\PZ$ and
  $\PZ\PZ$ production in the POWHEG-BOX-V2}'',} \textit{ Eur. Phys. J. C}
  \textbf{ 74} (2014) 2702,
  \href{http://dx.doi.org/10.1140/epjc/s10052-013-2702-5}{\doi{10.1140/epjc/s10052-013-2702-5}},
  \href{http://www.arXiv.org/abs/1311.1365}{\texttt{arXiv:1311.1365}}.

\bibitem{Alwall:2014hca}
J.~Alwall\hrefCMSnoop {}{ {et~al.}, ``{The automated computation of tree-level
  and next-to-leading order differential cross sections, and their matching to
  parton shower simulations}'',} \textit{ JHEP} \textbf{ 07} (2014) 079,
  \href{http://dx.doi.org/10.1007/JHEP07(2014)079}{\doi{10.1007/JHEP07(2014)079}},
  \href{http://www.arXiv.org/abs/1405.0301}{\texttt{arXiv:1405.0301}}.

\bibitem{Frederix:2012ps}
\hrefCMSnoop {}{R.~Frederix and S.~Frixione, ``{Merging meets matching in
  MC@NLO}'',} \textit{ JHEP} \textbf{ 12} (2012) 061,
  \href{http://dx.doi.org/10.1007/JHEP12(2012)061}{\doi{10.1007/JHEP12(2012)061}},
\href{http://www.arXiv.org/abs/1209.6215}{\texttt{arXiv:1209.6215}}.

\bibitem{Sjostrand:2014zea}
T.~Sj{\"o}strand\hrefCMSnoop {}{ {et~al.}, ``{An introduction to PYTHIA
  8.2}'',} \textit{ Comput. Phys. Commun.} \textbf{ 191} (2015) 159,
  \href{http://dx.doi.org/10.1016/j.cpc.2015.01.024}{\doi{10.1016/j.cpc.2015.01.024}},
\href{http://www.arXiv.org/abs/1410.3012}{\texttt{arXiv:1410.3012}}.

\bibitem{Sirunyan:2019dfx}
\hrefCMSnoop {}{{CMS Collaboration}, ``{Extraction and validation of a new set
  of CMS PYTHIA8 tunes from underlying-event measurements}'',} \textit{ Eur.
  Phys. J. C} \textbf{ 80} (2020) 4,
  \href{http://dx.doi.org/10.1140/epjc/s10052-019-7499-4}{\doi{10.1140/epjc/s10052-019-7499-4}},
  \href{http://www.arXiv.org/abs/1903.12179}{\texttt{arXiv:1903.12179}}.

\bibitem{NNPDF:2017mvq}
\hrefCMSnoop {}{{NNPDF} Collaboration, ``{Parton distributions from
  high-precision collider data}'',} \textit{ Eur. Phys. J. C} \textbf{ 77}
  (2017) 663,
  \href{http://dx.doi.org/10.1140/epjc/s10052-017-5199-5}{\doi{10.1140/epjc/s10052-017-5199-5}},
  \href{http://www.arXiv.org/abs/1706.00428}{\texttt{arXiv:1706.00428}}.

\bibitem{Agostinelli2003250}
\hrefCMSnoop {}{{{\GEANTfour}} Collaboration, ``{{\GEANTfour}---a simulation
  toolkit}'',} \textit{ Nucl. Instrum. Meth. A} \textbf{ 506} (2003) 250,
\href{http://dx.doi.org/10.1016/S0168-9002(03)01368-8}{\doi{10.1016/S0168-9002(03)01368-8}}.

\bibitem{Oreglia:1980cs}
\href {http://www.slac.stanford.edu/pubs/slacreports/slac-r-236.html}{M.~J.
  Oreglia, ``A study of the reactions $\psi^\prime \to \gamma \gamma \psi$''}.
\newblock PhD thesis, Stanford University, 1980.
\newblock {{SLAC} Report {SLAC-R-236}, see Appendix {D}}.

\bibitem{Gaiser:1982yw}
\href
  {http://slac.stanford.edu/pubs/slacreports/reports02/slac-r-255.pdf}{J.~E.
  Gaiser, ``{Charmonium spectroscopy from radiative decays of the $J/\psi$ and
  $\psi^\prime$}''}.
\newblock PhD thesis, Stanford University, 1982.
\newblock {{SLAC} Report {SLAC-R-255}}.

\bibitem{Dauncey:2014xga}
\hrefCMSnoop {}{P.~D. Dauncey, M.~Kenzie, N.~Wardle, and G.~J. Davies,
  ``{Handling uncertainties in background shapes}: {the discrete profiling
  method}'',} \textit{ JINST} \textbf{ 10} (2015) P04015,
  \href{http://dx.doi.org/10.1088/1748-0221/10/04/P04015}{\doi{10.1088/1748-0221/10/04/P04015}},
  \href{http://www.arXiv.org/abs/1408.6865}{\texttt{arXiv:1408.6865}}.

\bibitem{ftest}
\hrefCMSnoop {}{R.~A. Fisher, ``On the interpretation of $\chi^{2}$ from
  contingency tables, and the calculation of {P}'',} \textit{ J. R. Stat. Soc.}
  \textbf{ 85} (1922) 87,
  \href{http://dx.doi.org/10.1098/rsta.1922.0009}{\doi{10.1098/rsta.1922.0009}}.

\bibitem{CMS-LUM-17-003}
\hrefCMSnoop {}{{CMS Collaboration}, ``{Precision luminosity measurement in
  proton-proton collisions at $\sqrt{s} = 13\TeV$ in 2015 and 2016 at CMS}'',}
  \textit{ Eur. Phys. J. C} \textbf{ 81} (2021) 800,
  \href{http://dx.doi.org/10.1140/epjc/s10052-021-09538-2}{\doi{10.1140/epjc/s10052-021-09538-2}},
  \href{http://www.arXiv.org/abs/2104.01927}{\texttt{arXiv:2104.01927}}.

\bibitem{CMS-PAS-LUM-17-004}
\href {https://cds.cern.ch/record/2621960}{{CMS Collaboration}, ``{CMS
  luminosity measurement for the 2017 data-taking period at $\sqrt{s} =
  13\TeV$}'',} {CMS Physics Analysis Summary} {CMS-PAS-LUM-17-004}, 2018.

\bibitem{CMS-PAS-LUM-18-002}
\href {https://cds.cern.ch/record/2676164}{{CMS Collaboration}, ``{CMS
  luminosity measurement for the 2018 data-taking period at $\sqrt{s} =
  13\TeV$}'',} {CMS Physics Analysis Summary} {CMS-PAS-LUM-18-002}, 2019.

\bibitem{Junk:1999kv}
\hrefCMSnoop {}{T.~Junk, ``{Confidence level computation for combining searches
  with small statistics}'',} \textit{ Nucl. Instrum. Meth. A} \textbf{ 434}
  (1999) 435,
  \href{http://dx.doi.org/10.1016/S0168-9002(99)00498-2}{\doi{10.1016/S0168-9002(99)00498-2}},
\href{http://www.arXiv.org/abs/hep-ex/9902006}{\texttt{arXiv:hep-ex/9902006}}.

\bibitem{Read:2002hq}
\hrefCMSnoop {}{A.~L. Read, ``{Presentation of search results: The $\CLs$
  technique}'',} \textit{ J. Phys. G} \textbf{ 28} (2002) 2693,
\href{http://dx.doi.org/10.1088/0954-3899/28/10/313}{\doi{10.1088/0954-3899/28/10/313}}.

\bibitem{Cowan:2010js}
\hrefCMSnoop {}{G.~Cowan, K.~Cranmer, E.~Gross, and O.~Vitells, ``Asymptotic
  formulae for likelihood-based tests of new physics'',} \textit{ Eur. Phys. J.
  C} \textbf{ 71} (2011) 1554,
  \href{http://dx.doi.org/10.1140/epjc/s10052-011-1554-0}{\doi{10.1140/epjc/s10052-011-1554-0}},
  \href{http://www.arXiv.org/abs/1007.1727}{\texttt{arXiv:1007.1727}}.
[Erratum: \DOI{10.1140/epjc/s10052-013-2501-z}].

\bibitem{ATLAS:2011tau}
\href {https://cdsweb.cern.ch/record/1379837}{{ATLAS and CMS Collaborations},
  ``Procedure for the {LHC H}iggs boson search combination in summer 2011'',}
  ATL-PHYS-PUB-2011-011, {CMS NOTE-2011/005}, 2011.

\end{thebibliography}\endgroup
